\documentclass[pdflatex,sn-basic]{sn-jnl}

\usepackage{graphicx}%
\usepackage{multirow}%
\usepackage{amsmath,amssymb,amsfonts}%
\usepackage{amsthm}%
\usepackage{mathrsfs}%
\usepackage[title]{appendix}%
\usepackage{xcolor}%
\usepackage{textcomp}%
\usepackage{manyfoot}%
\usepackage{booktabs}%
\usepackage{algorithm}%
\usepackage{algorithmicx}%
\usepackage{algpseudocode}%
\usepackage{listings}%
\usepackage{multicol}
\usepackage{gensymb}
\usepackage{geometry}
\usepackage{bm}
\usepackage{comment}
\usepackage{subcaption}
\usepackage{xurl}
\usepackage{placeins}

\usepackage[export]{adjustbox}

\makeatletter
\@twosidefalse
\@mparswitchfalse
\makeatother

\DeclareMathOperator*{\argmax}{arg\,max} 
\newcommand{\fil}[1]{\widehat{#1}}
\newcommand{\dt}[1]{\frac{\partial#1}{\partial t}}
\newcommand{\dxi}[1]{\frac{\partial#1}{\partial x_i}}
\newcommand{\dxj}[1]{\frac{\partial#1}{\partial x_j}}
\newcommand{\dxjj}[1]{\frac{\partial^2#1}{\partial x_j \partial x_j}}

\theoremstyle{thmstyleone}%
\theoremstyle{thmstyletwo}%

\theoremstyle{thmstylethree}%

\begin{document}

\title[WindRL]{Reinforcement Learning Increases Wind Farm Power Production by Enabling Closed-Loop Collaborative Control}

\author*[1]{\fnm{Andrew} \sur{Mole}}\email{a.mole@imperial.ac.uk}
\author[1]{\fnm{Max} \sur{Weissenbacher}}\email{m.weissenbacher@imperial.ac.uk}
\author[1]{\fnm{Georgios} \sur{Rigas}}\email{g.rigas@imperial.ac.uk}
\author[1]{\fnm{Sylvain} \sur{Laizet}}\email{s.laizet@imperial.ac.uk}

\affil*[1]{\orgdiv{Department of Aeronautics}, \orgname{Imperial College London}, \orgaddress{ \city{London}, \country{UK}}}

\abstract{

Traditional wind farm control operates each turbine independently to maximize individual power output.
However, coordinated wake steering across the entire farm can substantially increase the combined wind farm energy production.
Although dynamic closed-loop control has proven effective in flow control applications, wind farm optimization has relied primarily on static, low-fidelity simulators that ignore critical turbulent flow dynamics.
In this work, we present the first reinforcement learning (RL) controller integrated directly with high-fidelity large-eddy simulation (LES), enabling real-time response to atmospheric turbulence through collaborative, dynamic control strategies.
Our RL controller achieves a $4.30\%$ increase in wind farm power output compared to baseline operation, nearly doubling the $2.19\%$ gain from static optimal yaw control obtained through Bayesian optimization.
These results establish dynamic flow-responsive control as a transformative approach to wind farm optimization, with direct implications for accelerating renewable energy deployment to net-zero targets.
}


\maketitle

\section{Introduction}\label{intro}

Wind energy is expected to play a crucial role in expanding renewable energy generation and decarbonizing energy production~\cite{IEA2021}.
Maximizing the energy output of existing wind farms through wind farm control is a key strategy to enhance the efficiency of renewable energy systems.
Modern large–scale offshore wind farms consist of multiple turbines grouped together, usually in well-structured formations.
The wind turbines are typically operated with a greedy strategy, optimizing their individual power generation whilst ignoring the effect on other wind turbines.
This leads to individual wind turbines often operating in suboptimal conditions due to aerodynamic interactions between the wind turbines, resulting in a reduction of the total power output of the wind farm~\citep{barthelmieEvaluationWindFarm2010}.
By instead implementing a collective wind farm control approach, that accounts for aerodynamic interactions between turbines, wake effects can be mitigated, leading to improved wind farm efficiency~\citep{howlandCollectiveWindFarm2022, harrison-atlasArtificialIntelligenceaidedWind2024a}.

Efficient control of wind farms remains one of the major challenges in wind energy science~\citep{veersGrandChallengesDesign2023} and has been the focus of extensive research.
Control strategies fall broadly into two categories: \emph{static} (or \emph{quasi-static}), where fixed optimal parameters such as turbine positions are obtained, and \emph{dynamic}, where parameters such as the pitch angle of the turbine blade change as a function of time.
Dynamic wind farm control can be further divided into \emph{open-loop control}, where control strategies follow a predetermined dynamic behavior, and \emph{closed-loop control}, where the controller senses and actively responds to changing flow conditions.
Various models are used to simulate wind farms, ranging from lower-fidelity wake models~\citep{jensen1983, jimenezApplicationTechniqueCharacterize2010a, bastankhahNewAnalyticalModel2014} to higher-fidelity LES solvers~\citep{calafLargeEddySimulation2010, meyersOptimalTurbineSpacing2012}.
Analytical wake models are computationally efficient but sacrifice accuracy, particularly in capturing transient flow dynamics and wake-to-wake interactions.
The most commonly studied control methods for wind farm optimization are axial induction control and yaw control, with yaw control generally showing more promising improvements in power generation~\citep{meyersWindFarmFlow2022a}.
Numerous studies have focused on identifying static optimal settings for these actuations, using a variety of optimization techniques~\citep{howlandCollectiveWindFarm2022, moleMultifidelityBayesianOptimisation2024, houckReviewWakeManagementTechniquies2022}, including surrogate-based Bayesian optimization.
A smaller body of work has explored open-loop dynamic control through periodic forcing, particularly for induction control~\cite{goitOptimalControlEnergy2015, muntersDynamicStrategiesYaw2018, frederikHelixApproachUsing2020}, and to a lesser extent for yaw control~\cite{muntersDynamicStrategiesYaw2018, howlandOptimalClosedloopWake2020}.
However, these studies have generally found that open-loop dynamic forcing provides limited or no improvement compared to quasi-static or static control strategies.
We refer the reader to the review by~\citet{meyersWindFarmFlow2022a} for an in-depth review of wind farm control studies.

Data-driven methods have emerged as a competitive alternative to classical control frameworks due to their ability to adapt to complex system dynamics as data become available.
Reinforcement Learning (RL) is a machine learning paradigm in which a controller (or agent) learns to make sequential decisions by interacting with an environment and receiving rewards based on its actions~\citep{sutton2018reinforcement}.
Well known for achieving superhuman performance in video games~\citep{mnih2015human, mnih2016asynchronous}, RL has since been successfully applied to flow control applications such as the reduction of drag in a 3D turbulent flow~\citep{chatzimanolakis2024learning} and the suppression of vortex shedding in partially observable environments~\citep{xia2023active}.
We highlight the recent work by~\citet{font2025deep}, who couple an RL control framework with a high-fidelity 3D flow simulation of a turbulent separation bubble.
They demonstrate superior performance of RL compared to classical open-loop periodic forcing control.
We refer the reader to the recent review by~\citet{vignon2023recent} for an overview of flow control with RL.

RL has been explored for the optimization of wind farms, but most studies have been limited to static control policies.
The majority of existing approaches~\citep{buiDistributedOperationWind2020, heEnsemblebasedDeepReinforcement2022, padullaparthiFALCONFArmLevel2022, dongIntelligentWindFarm2021, monrocWFCRLMultiAgentReinforcement2025, wangDeepReinforcementLearningdriven2025} rely on analytical wake models that capture only the steady-state, time-averaged flow field.
These models neglect the transient dynamics of the atmospheric boundary layer and wake interactions, making it impossible to learn dynamic closed-loop controllers.
Studies that combine RL with high-fidelity simulations are rare. \citet{korbExploringApplicationReinforcement2021} couple a PPO~\citep{schulman2017proximal} RL controller with a 3D LES simulation of three in-line turbines, controlling both generator torque and fixed-frequency blade pitch oscillations.
However, neither approach results in an effective dynamic closed-loop strategy with the torque control failing to increase power output, while pitch control converges to a nearly static amplitude, effectively converging to an open-loop forcing controller.
To date, no studies have successfully demonstrated an RL controller that achieves efficient, dynamic, closed-loop wake steering based on real-time flow observations in a high-fidelity simulation environment.

In the present work, we combine an RL control framework with high-fidelity large-eddy simulations (LES) of a small three-turbine wind farm.
We obtain a novel closed-loop controller that dynamically adapts the turbine yaw angles to instantaneous wind conditions around the farm.
To the best of our knowledge, this is the first successful implementation of an efficient dynamic closed-loop RL active flow controller for wind farm control.

\section{Results}\label{sec:results}

\subsection{RL Training and Bayesian Optimization}
\begin{figure}[h]
    \centering
    \begin{minipage}{0.45\textwidth}
        \centering
        \includegraphics[width=0.8\linewidth]{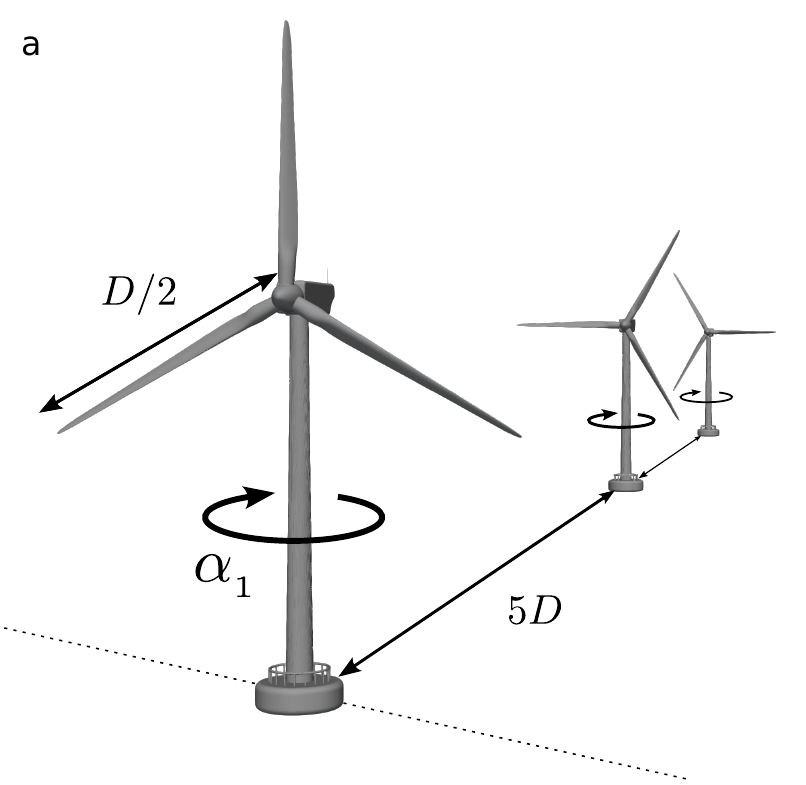}
        \vspace{0.75cm}
        \includegraphics[width=\linewidth]{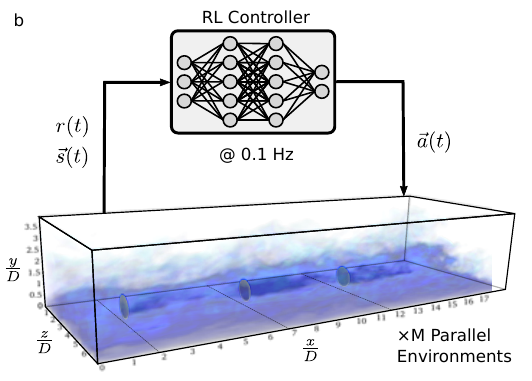}

    \end{minipage}
    \begin{minipage}{0.5\textwidth}
        \vfill
        \includegraphics[width=\linewidth]{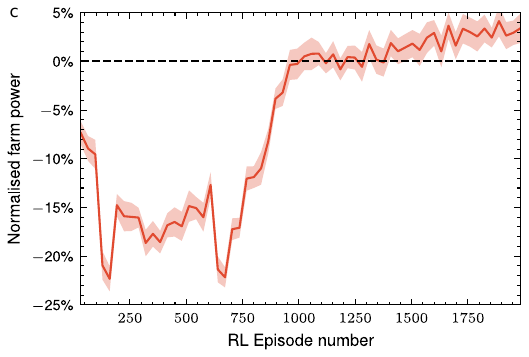}
        \vspace{0.5cm}
        \includegraphics[width=\linewidth]{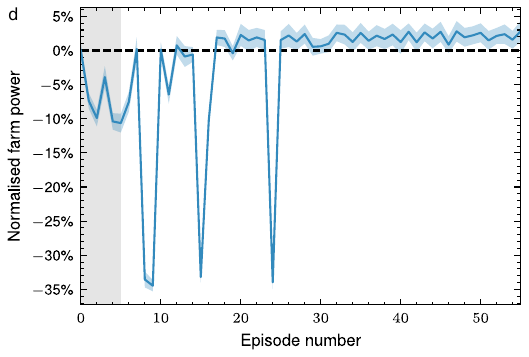}
    \end{minipage}
    \caption{\emph{(a)}: Schematic of the simulated wind farm. Three in-line turbines with diameter $D=126m$ in an atmospheric boundary layer are simulated using a high-fidelity 3D-LES solver. The turbines are aligned along the direction of the flow and the yaw angle is dynamically adjusted. \emph{(b)}: Schematic of the RL control loop. The controller receives velocity sensor observations from $M=32$ LES environments and determines the yaw angles of three in-line turbines. Angles are updated by the controller every $10s$ of simulated time. \emph{(c)}: Mean farm power during RL training relative to the greedy baseline with $95\%$ CI computed over $n=32$ training environments as a function of episode number. \emph{(d)}: Mean relative farm power with $95\%$ CI computed over $n=32$ training environments during Bayesian optimization to find optimal static yaw angles.}
    \label{fig:rl_training}
\end{figure}

A set of baselines is established in order to fairly assess the performance of the discovered RL control strategies.
In standard wind farm layouts, individual turbines face directly into the average oncoming wind, which in our case of three in-line wind turbines (as in Figure~\ref{fig:rl_training}(a)) corresponds to $\vec{\alpha}_{\mathrm{greedy}} = (0\degree, 0\degree, 0\degree)$.
We refer to this as the `greedy baseline'. In addition, we use a Bayesian optimizer (see Methods section) to compute the optimum \emph{static} angles of individual turbines.
The optimization progress is shown in Figure~\ref{fig:rl_training}(d), initially exhibiting substantial fluctuations with peaks below $-30\%$ farm power, reflecting the exploration as Bayesian Optimization (BO) samples broadly across the parameter space.
As optimization progresses, the resulting wind farm power consistently exceeds the greedy baseline.
The optimal (static) turbine yaw angles found in the BO are $\vec{\alpha}_{\mathrm{BO}} = (20\degree, 13\degree, -3\degree)$.
These angles show a decrease from the upstream to the downstream turbines, which is consistent with established wake steering strategies~\citep{houckReviewWakeManagementTechniquies2022}, where optimal power production is achieved when the upstream turbines are yawed more significantly, with reduced yaw angles for the downstream turbines.

In order to ensure effective RL convergence, we conduct a set of hyperparameter sweeps.
We find that using a larger number of velocity sensors located in the vicinity of the wind farm is advantageous, and that carefully tuning the entropy learning rate (a hyperparameter that controls the degree of exploration by the agent as training progresses by encouraging randomness in its policy) is imperative to prevent premature convergence to a local optimum.
We train our RL controller using the selected hyperparameters (see Appendix) for a total of $1 \times 10^6$ interactions collected across $32$ parallel LES environments, see Figure~\ref{fig:rl_training}(b).
The progress of the RL training is shown in Figure~\ref{fig:rl_training}(c) with the RL control initially reducing farm power output, including two pronounced troughs below $–20\%$, as the agent explores the policy space driven by the regularization of the entropy of the objective.
After approximately 1,000 episodes, the policy consistently increases the wind farm power output over the greedy baseline, and continues to improve the performance gradually as it transitions from broad exploration to more focused exploitation of the policy.
Training the RL controller in the simulated environments required $18.5h$ on $34$ dual AMD EPYC 7742 64-core processor nodes on the ARCHER2 system~\citep{archer2}.
Energy consumption was recorded using ARCHER2’s job-level energy accounting tools and totaled approximately $254 kWh$ per training run.

\subsection{Mean Farm Power Increase}

We evaluated the RL controller, the static Bayesian optimum, and the greedy baseline by collecting a total of $n=224$ episodes across $16$ parallel LES environments with independent inlet and atmospheric boundary conditions for each case.
Each episode lasts $2 \times 10^4 s$ and the power signal is sampled at $5 \, \mathrm{Hz}$.
The mean farm power $\bar{P}$ is calculated by averaging the instantaneous power over the duration of an episode.

\begin{figure}[h]
    \includegraphics[width=\linewidth]{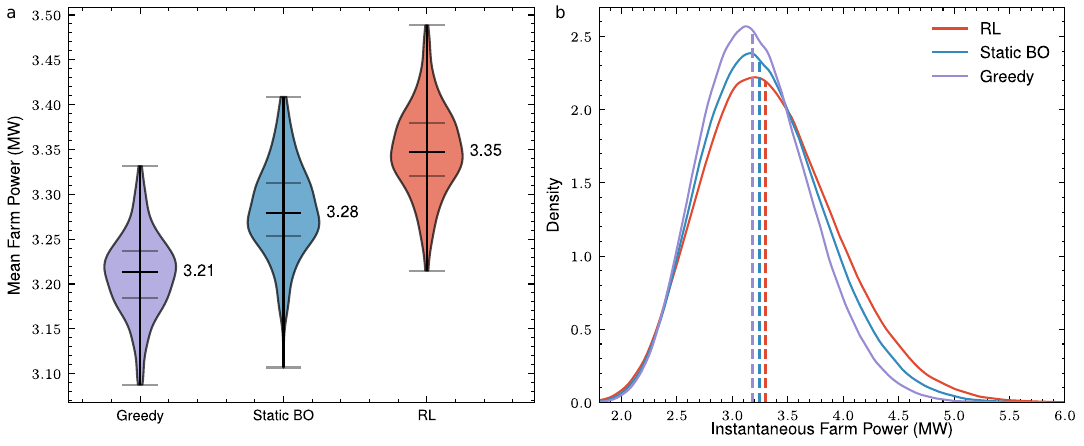}
    \caption{\emph{(a)}: Violin plot of the mean farm power in MW computed over $n=224$ evaluation episodes across $16$ environments. The horizontal lines are (bottom to top): minimum, first quantile, median, third quantile, maximum. The annotated value is the median. The active RL controller outperforms both the greedy case and the static Bayesian optimum. \emph{(b)}: Density of the \emph{instantaneous} power output normalized so that the integral of each curve is the mean farm power. The dashed vertical lines indicate the medians of the distributions. The RL controller achieves a higher mean power by spending more time in higher power output conditions.}
    \label{fig:mean_power}
\end{figure}

Figure~\ref{fig:mean_power}(a) shows the distribution of the mean farm power for each case.
Relative to the greedy baseline, the static Bayesian optimum increases the mean farm power output by $2.19\%$ (95\% confidence intervals (CI) = $[1.98\%, 2.39\%]$), while the RL controller increases the mean power by $4.30\%$ (95\% CI = $[4.10\%, 4.49\%]$).
The magnitude of the power increase of each strategy depends on the atmospheric boundary layer conditions as well as the wind turbine specifications and layout.

The RL controller robustly increases the mean farm power while outperforming both the greedy baseline and the static Bayesian optimum.
The higher mean power is achieved by spending more time occupying higher power generation states.
This can be seen in the plot of the instantaneous power output distributions in Figure~\ref{fig:mean_power}(b).
The peak of the distribution is reduced relative to the greedy case, showing a more variable power output with a higher distribution across the high-power tail of the distribution.
A similar, though less pronounced, shift is observed for the static Bayesian optimization case, in line with its more modest power increase.

\subsection{Closed-Loop Controller Test}\label{subsec:closed}
We verify that the trained RL controller is performing active \emph{closed-looped} control as opposed to having learned an open-loop forcing strategy. To accomplish this, we evaluate the agent on 16 environments and save the sequence of actions performed. We then `replay' this sequence of actions on 16 environments with \emph{statistically independent} inlet conditions. We collect one episode containing $2 \times 10^4 s$ per environment ($n=16$ samples in total) and compute the mean power $\bar{P}$ for each. In this test, the `replayed' controller does not result in a statistically significant increase in power (95\% CI = $[-0.35\%, +0.96\%]$) compared to the greedy case. We conclude that, because accurate sensor measurements are required for the RL controller to find the power improvements, the learned RL controller is closed-loop; it actively responds to the changing flow conditions instead of forcing the flow at a certain fixed frequency.

\subsection{Dynamic switching}

\begin{figure}[!h]
    \centering
    \includegraphics[width=\linewidth]{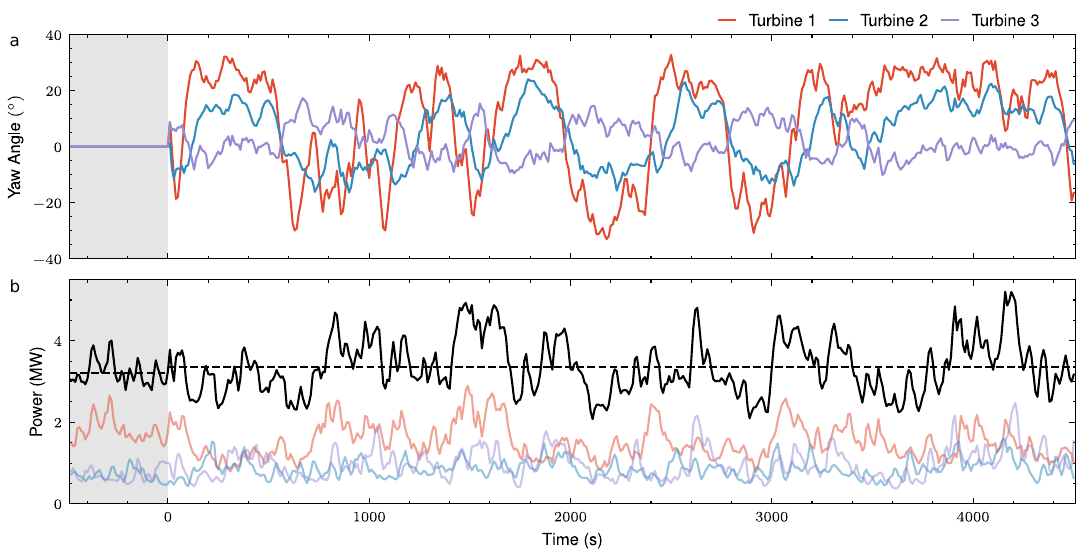}
    \includegraphics[width=\linewidth]{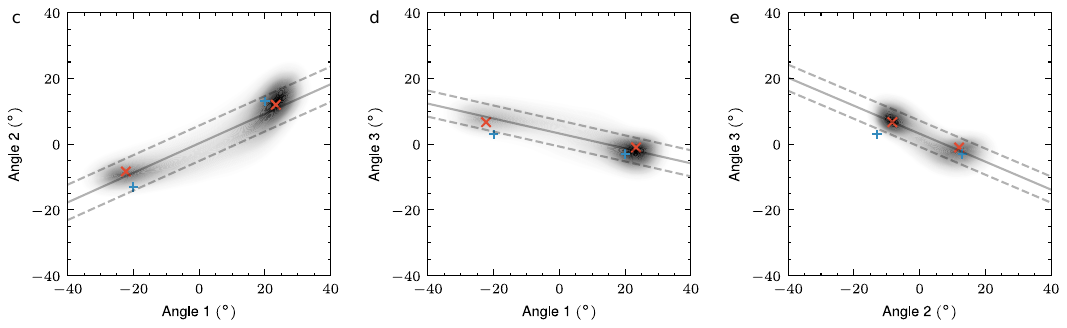}
    \includegraphics[width=\linewidth]{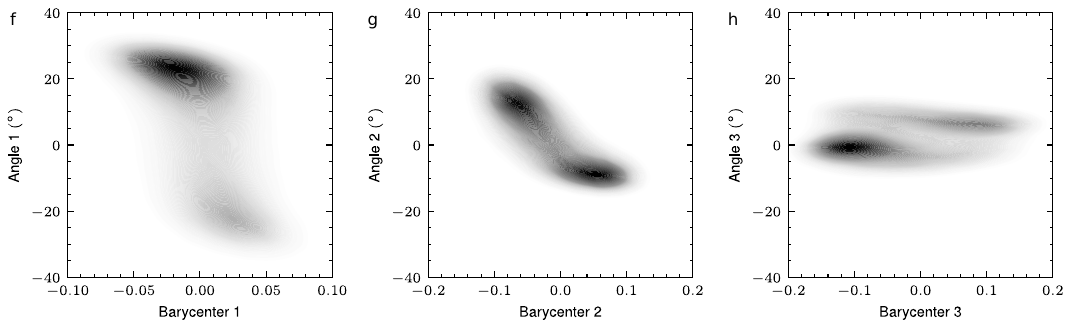}
    \caption{\emph{(a)} Turbine yaw angles and \emph{(b)} power output for an evaluation episode of the converged RL controller. The controller starts at time $t=0s$, and the mean power (dotted horizontal black line) increases. The black line in the power plot is the total wind farm power (the sum of the individual turbine powers). \emph{(c)(d)(e)}: Joint densities of angle pairs after correcting for time shifts. Time shifts are found using cross-correlation analysis. Linear models are fitted using least squares error. Red $\times$'s show the peaks of the RL densities $\vec{\alpha}_{\mathrm{RL}}$ and the blue $+$'s show the Bayesian static optimum locations and their symmetric inflections $\vec{\alpha}_{\mathrm{BO}}$. \emph{(f)(g)(h)}: Joint densities between the angle of each turbine and the barycenter of the velocity in front of that turbine.}
    \label{fig:controller}
\end{figure}

Figure~\ref{fig:controller}(a-b) shows a time series segment of an evaluation rollout of the RL controller, showing the yaw angles and corresponding power outputs. 
Initially, all yaw angles are at $0^{\circ}$, corresponding to the pre-control baseline.
When the controller is activated, the turbines' yaw angles are dynamically adjusted, following the learned policy, with a combination of oscillations and discrete switching between yaw states.
The resulting power output signals show strong fluctuations, driven in part by variability in the atmospheric boundary layer.
The mean power generation improvement exhibited by the RL controller is obscured by these fluctuations.
Consistently with established wake steering strategies, the controller reduces the power generation of the first turbine, which is offset by the increased energy capture of the downstream turbines.
Although the time series illustrates the dynamic behavior of the RL controller during a single rollout, to better understand its behavior, statistical characterizations are drawn from the $n=224$ evaluation episodes.
These distributions reveal consistent control patterns and performance trends that are not apparent from individual trajectories alone.

Firstly, a cross-correlation analysis of the turbine yaw angles (see more details in the Appendix) is conducted, showing distinct peaks at specific time delays.
Relative to the angle of the first turbine, the angle of the second turbine is best correlated with a delay of $+70s$.
This is approximately the advection time based on the free-stream flow ($U_0 = 7.5 ms^{-1}$) from one turbine to the next ($\Delta t = 5 D /U_{\infty} = 84s$).
The RL controller has therefore learned that there is an inherent time delay in the system, created by the finite speed of propagation of the fluid flow.
Interestingly, the third turbine correlation peak has a delay of $-30s$ relative to the second, meaning that the third turbine acts before the second, a fact the authors attribute to the fact that the propagation time between these turbines is not an important feature for power extraction and the third turbine has no downstream turbines to account for.

After compensating for time lags, the turbines appear to move in synchronization.
Figure~\ref{fig:controller}(c)(d)(e) shows the joint densities of the yaw angle signals when adjusting for the time delays, revealing a strong positive correlation between the first and second turbines and a negative correlation between the rear turbine and the others.
This hypothesis is verified by fitting linear models to the yaw angle pairs.
 Good agreement is found, with $R^2 = 0.7114$ for the front pair ($\alpha_2$ vs.\ $\alpha_1$), and $R^2 = 0.5370$ and $R^2 = 0.5358$ for the rear turbine's relationships to $\alpha_1$ and $\alpha_2$, respectively and all mean residuals below $4.3^\circ$.
 The details of the fitted linear models are given in the Appendix.

Closer inspection of the density plot in Figure~\ref{fig:controller} reveals that the yaw angles tend to concentrate around two local modes $\vec{\alpha}_{\mathrm{RL}}^+ = (23 \degree, 12\degree, -1\degree)$ respectively $\vec{\alpha}_{\mathrm{RL}}^- = (-22\degree, -8\degree, 7\degree)$ (rounded to nearest integer).
The modes are close in absolute value to the Bayesian static optimum $\vec{\alpha}_{\mathrm{BO}} = (20\degree, 13\degree, -3\degree)$ and its symmetric inflection $-\vec{\alpha}_{\mathrm{BO}}$, respectively.
Furthermore, when using $\vec{\alpha}_{\mathrm{RL}}^{\pm}$ as static angles, the mean power generated is nearly identical to the power generated by the static optimum $\vec{\alpha}_{\mathrm{BO}}$: When used as static angle configurations, $\vec{\alpha}_{\mathrm{RL}}^+$ generates $-1.36\%$ and $\vec{\alpha}_{\mathrm{RL}}^-$ generates $-1.59 \%$ less power than the optimal static angle $\vec{\alpha}_{\mathrm{BO}}$. Since the mean flow through the turbines is statistically symmetric in the stream-wise direction, it is expected that $\vec{\alpha}$ and $-\vec{\alpha}$ generate the same power.
However, this symmetry holds only in a long-time average sense, with the instantaneous flow field being highly asymmetric due to unsteady phenomena, such as turbulent gusts, breaking this symmetry at any given moment.

We hypothesize that the RL controller has learned to exploit the instantaneous asymmetries by dynamically switching between the two symmetric static optima $\vec{\alpha}_{\mathrm{BO}}$ and $-\vec{\alpha}_{\mathrm{BO}}$.
In fact, the strategy switches between $\vec{\alpha}_{\mathrm{RL}}^+$ and $\vec{\alpha}_{\mathrm{RL}}^-$, which when used as static angles produce a mean power very close to that of the Bayesian optima.

Further insight into the controller’s sensing strategy is provided by examining the joint density plots between the yaw angle of each turbine and the lateral position of the velocity barycenter directly upstream (Figure~\ref{fig:controller}, bottom row).
For each turbine $t$, the barycenter $\left<z\right>_t$ is computed from the stream-wise velocity measured at fourteen probes located $1D$ and $1.5D$ upstream of each turbine at hub height.
This is a subset of the signals used as observations by the RL controller.
The discrete barycenter is given by:
\begin{equation}
    \left<z\right>_t = \frac{\sum_{i=1}^{14} u_i(z-z_t)}{D\sum_{i=1}^{14} u_i} \; ,
\end{equation} 
where $u_i$ is the stream-wise velocity measured at lateral position $z_i$.
For the first and second turbines, a clear negative correlation is observed between the barycenter and the yaw angle (see Figure~\ref{fig:controller}(f)(g)(h)) meaning that when the incoming flow shifts laterally in one direction, the RL controller responds by yawing the turbine to deflect the flow in the opposite direction.
This indicates that the controller exhibits a reactive steering mechanism for which the turbines counteract the incoming asymmetry to redirect higher-momentum flow toward the downstream turbines.
For the third turbine, the correlation is less clear, possibly because the incoming flow is more intermittent as a result of wake-to-wake interactions.
The control of the third turbine does not need to balance its power output with its effect on the other turbines (as there are no turbines downstream of it), which explains its different response to the flow fluctuations in order to maximize its own power extraction.
Overall, these correlations suggest that the RL controller has not only learned time-coordinated behaviors but also developed a spatially aware policy that maps upstream flow features to turbine actuation in a responsive, feedback-driven manner.

\subsection{Spectral densities and Bandwidth-limiting analysis}
\begin{figure}[!h]
    \centering
        \includegraphics[width=\linewidth]{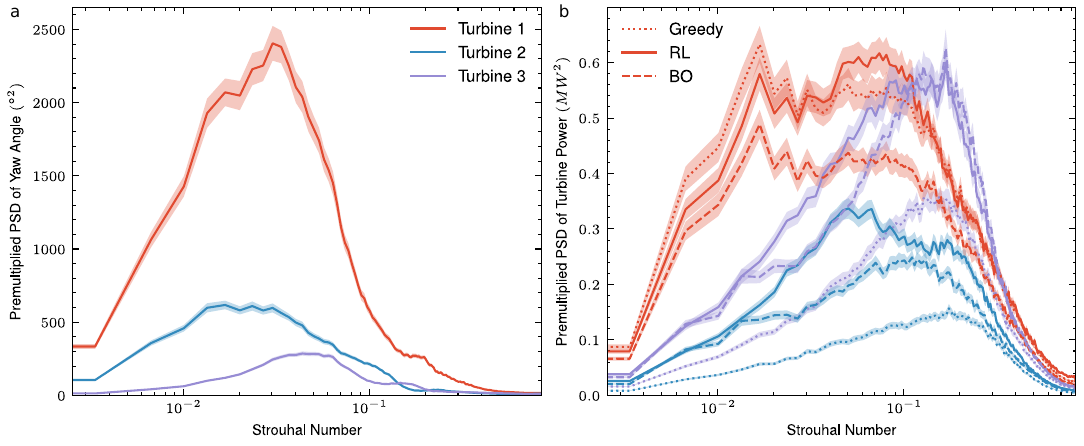}
        \includegraphics[width=\linewidth]{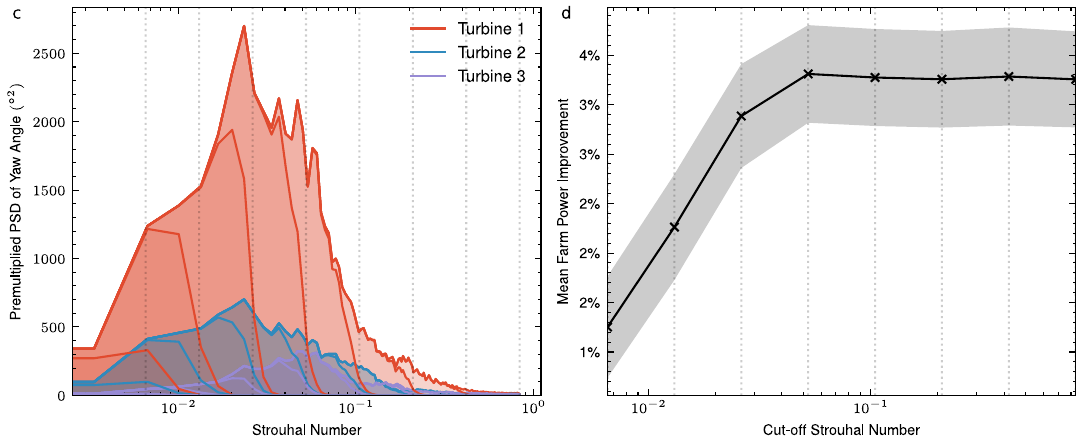}
        \caption{\emph{(a)}: Premultiplied power spectral density of turbine yaw angles of the controller. \emph{(b)}: Premultiplied power spectral density of the individual turbine powers for the uncontrolled (greedy) case (dotted), the static BO case (dashed) and the dynamic controlled (RL) case (solid). Shaded areas show the $95\%$ CI computed over $n=224$ evaluation environments. \emph{(c)}: Power spectral density of turbine yaw angles with various low-pass filters applied shown by dotted vertical lines. \emph{(d)}: The corresponding mean farm power improvements for the different low-pass filtered controllers.}
    \label{fig:frequencies}
\end{figure}

A spectral analysis of the yaw angle signals obtained from evaluation reveals that the dominant frequency component is located at $\mathrm{St} \approx 2 \times 10^{-2}$, see Figure~\ref{fig:frequencies}(a).
Here, the Strouhal number $\mathrm{St} = f D / U_0$, with $D$ the turbine diameter, $U_0$ the free-stream velocity at turbine hub height and $f$ the measured frequency.
The yaw angle is actuated across a broad range of frequencies, suggesting that it is targeting turbulent features across a wide range of temporal scales.
The dominant actuation frequency is far from the wake meandering ($\mathrm{St} \approx 2 \times 10^{-1}$) and vortex shedding ($\mathrm{St} \approx 3 \times 10^{-1}$) frequencies.
The resulting frequency response of the three turbines for each of the greedy, static BO, and RL cases are shown in Figure~\ref{fig:frequencies}(b).
Compared to the static BO case, which reduces the strength of the power fluctuations of the first turbine compared to the greedy case, the RL case only marginally reduces the power fluctuations at low frequencies and marginally increases them in the range $0.05 \leq St \leq 0.1$.
The second and third turbines see greater increases in power fluctuation in the RL case than in the BO case.
This is likely due to the increase in the wind speeds and subsequent mean power output of these turbines.

To assess the relative importance of each frequency range, we conducted a bandwidth-limiting study.
We use Welch's method to low-pass filter the yaw angle signal at decreasing cut-off frequencies starting at half the Nyquist frequency; and then replay the filtered angle signal in the environments with inlet conditions \emph{identical} to the conditions used to compute the angle signals.
The result is that we are able to limit the effective frequency from the original $0.1 \, \mathrm{Hz}$ to $3.125 \times 10^{-3} \, \mathrm{Hz}$, see Figure~\ref{fig:frequencies}(c)(d).
The maximum angular yawing velocity experienced by the turbines is reduced by more than half from $1 \degree s^{-1}$ to below $0.4 \degree s^{-1}$.

\subsection{Flow analysis}

\begin{figure}[!h]
    \centering
    \includegraphics[width=\linewidth]{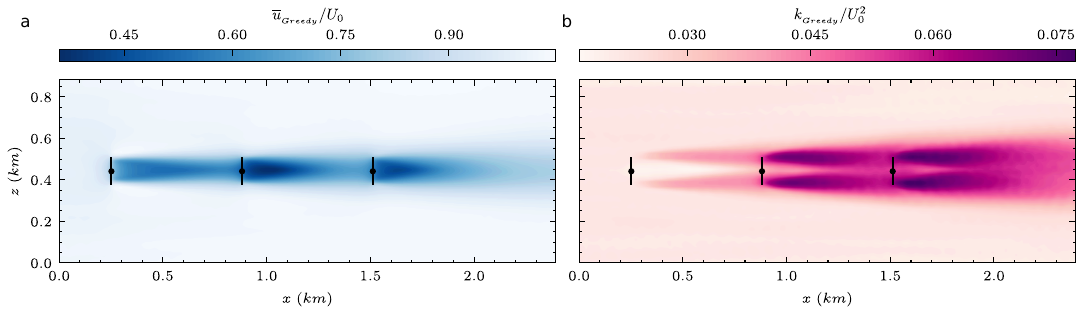}
    \includegraphics[width=\linewidth]{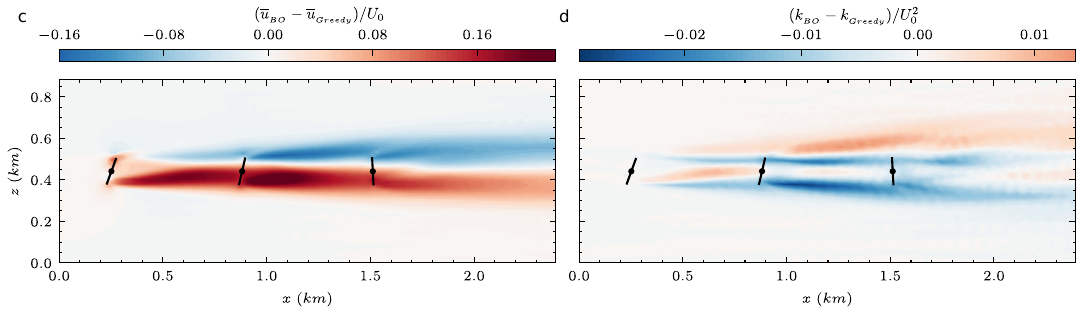}
    \includegraphics[width=\linewidth]{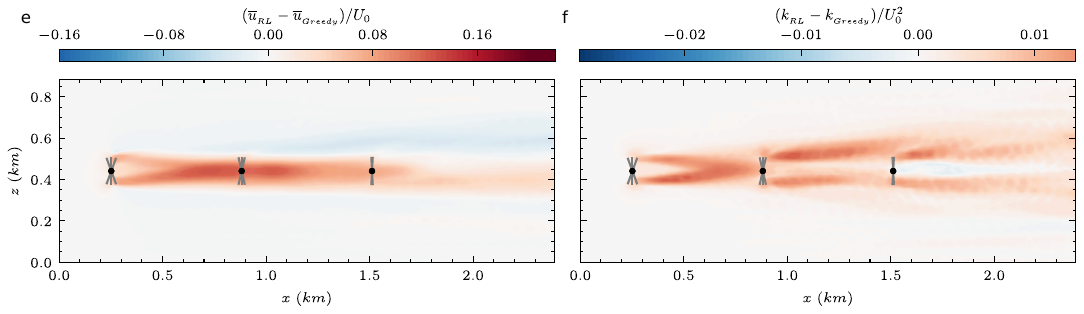}
    \caption{Slices of the flow field at the wind turbine hub height showing (a) the time-averaged stream-wise velocity and (b) the resolved turbulent kinetic energy of the flow around the turbines in the greedy case. The change in the (c) time-averaged stream-wise velocity and (d) the resolved turbulent kinetic energy, when the turbines are yawed at constant angles found by the BO. The  change in the (e) time-averaged stream-wise velocity and (f) the resolved turbulent kinetic energy, when the turbines are controlled actively by the RL controller.}
    \label{fig:flow_stats}
\end{figure}

To evaluate the effect of the control strategy on the flow field around the wind farm, we compare the time-averaged stream-wise velocity and resolved turbulent kinetic energy (TKE) fields on slices at the turbine hub height for the baseline greedy control, the static optimal configuration obtained via BO, and the RL controller (Figure~\ref{fig:flow_stats}).

In the greedy case (Figure~\ref{fig:flow_stats}(a)(b)), the reduced velocity of the wind behind the turbines aligns with the downstream turbine locations, which is the cause of the reduced power output of the second and third turbines.
The velocity gradients caused by the turbines result in an increase in the turbulent kinetic energy, particularly concentrated immediately behind the turbines.

When the turbines are statically yawed at the angles determined by the BO, the wakes are deflected laterally.
This results in both a relocation in the position of the wake and a partial recovery in flow speed.
The second and third turbines in this case experience an increase in mean velocity on one side and a decrease on the other relative to the greedy configuration (Figure~\ref{fig:flow_stats}(c)).
When averaged over each of these turbines swept area, the net effect is an increase in the wind speed and a corresponding increase in the power outputs of these turbines.
The turbulent kinetic energy shows an overall reduction in Figure~\ref{fig:flow_stats}(d), with localized increases near the edges of the deflected wakes.

In the RL-controlled case, the time-averaged velocity field around the three turbines shows reduced wake intensity and higher mean wind speeds compared to the baseline.
Notably, the increase in the averaged wind speed is concentrated at the center-lines of the second and third turbines, where power recovery is most critical.
Additionally, the RL controller induces a broad increase in the turbulent kinetic energy, in contrast to the redistribution observed with the static BO control.
This increased turbulence within the wake may contribute to improved mixing and accelerated wake recovery, contributing to the overall increase in power output achieved by the RL strategy.

\begin{figure}[!h]
    \centering
    \includegraphics[width=\linewidth]{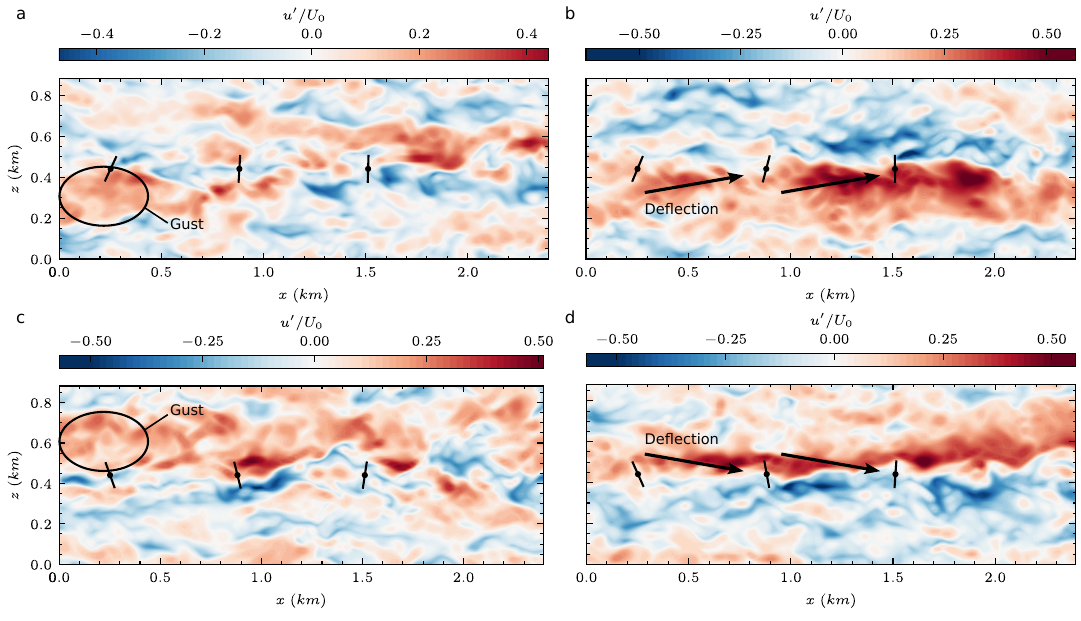}
    \caption{Slices of the flow field at the wind turbine hub height showing the instantaneous fluctuations of the stream-wise velocity at different times in the simulation (a) $t=1250s$ (b) $t=1500s$ (c) $t=2000s$ (d) $t=2250s$~. }
    \label{fig:flow_inst}
\end{figure}

To further highlight the RL controller's dynamic and adaptive behavior, we analyze instantaneous snapshots of the stream-wise velocity fluctuations $u' = u - \overline{u}$, at hub height across four time instances (Figure~\ref{fig:flow_inst}).
At $t = 1250,\text{s}$ (Figure~\ref{fig:flow_inst}a), a high-speed gust originating from the atmospheric boundary layer approaches the first-row turbine from the right side of the wind farm.
In response, the RL controller yaws the first turbine to deflect the gust toward the centerline, guiding the high-momentum flow toward the downstream turbines.
By $t = 1500,\text{s}$ (Figure~\ref{fig:flow_inst}b), the second turbine has also adjusted its orientation to continue steering the gust toward the third turbine, maximizing its beneficial impact.
The symmetric controller behavior is shown later in the simulation at $t = 2000,\text{s}$ (Figure~\ref{fig:flow_inst}c), a gust approaches from the right side, and again the upstream turbines yaw angles are adjusted to redirect the incoming higher momentum wind toward the center as seen at $t = 2250,\text{s}$ (Figure~\ref{fig:flow_inst}d).
This behavior demonstrates the RL controller's ability to sense asymmetries in the incoming flow and coordinate the turbine yaw angles in real time to opportunistically capture transient increases in available wind energy.
The switching behavior between two dominant yawing modes, previously characterized as $\vec{\alpha}_{\mathrm{RL}}^{+}$ and $\vec{\alpha}_{\mathrm{RL}}^{-}$, is reflected here in the gust directionality and the controller’s selection of the corresponding coordinated yaw pattern. This dynamic switching, combined with flow-aware coordination, underlies the ability of the RL controller to outperform static BO-derived yaw configurations.

\section{Discussion} \label{sec:conclusions}
In the present study, we explored the potential of collaborative dynamic wake steering to optimize the power output of a wind farm.
We coupled a high-fidelity LES of a small wind farm with an RL controller, allowing the controller to dynamically update the yaw angles of the turbines by sensing the velocity field around the turbines.
The converged RL controller yields a $+ 4.3\%$ increase in power over the greedy baseline, an improvement of almost a factor $2$ over the increase in power obtained with static optimal angles found using Bayesian optimization.

The learned control strategy is analyzed in detail.
First, we verified that the controller is indeed closed-loop, meaning that the strategy actively takes advantage of the sensor information.
By analyzing the cross-correlation between turbines, we showed that the controller dynamically switches between two almost-symmetric local optima, which are close to the optima found using static Bayesian optimization.
We confirm that this switching is based on the location of wind gusts formed in the atmospheric boundary layer.
A detailed analysis of the flow fields shows how the implementation of the controller changes the wind conditions around the turbines allowing for the increased power capture.

The learned controller operates at an effective frequency of $3 \times 10^{-3} \, \mathrm{Hz}$, with a maximum angular velocity of less than $0.4 \deg s^{-1}$, which is likely achievable for latest-generation commercial wind turbines. 
However, implementing such controllers in operational wind farms involves challenges related to computational demands, sensing requirements, and real-world transferability, in addition to an analysis of component fatigue.

One barrier is the computational cost of training.
Our RL agent requires thousands of episodes to converge, which is only feasible in simulation with parallelization of the environments.
Collecting this experience solely through real-world online training is impractical due to the slow dynamics of wind farms and the associated operational risks.
Consequently, deployment relies on training using high-fidelity simulated environments, with the trained policy then exported to run in real time.

Fortunately, once trained, the policy's neural network is computationally inexpensive to evaluate and can produce control decisions in real time.
RL has already shown success in other domains with similar or faster real-time constraints, such as robotics~\citep{koberReinforcementLearningRobotics2013} and energy systems~\citep{degraveMagneticControlTokamak2022}.
To enable the transfer from simulated training environment to real-world deployment, adequate sensing of the environment is required.
The RL wake steering requires anticipatory control, which depends on upstream wind field information, as shown in Section~\ref{subsec:closed}, which may be realized through turbine-mounted nacelle LiDAR systems, providing upstream velocity measurements at the required frequency.
This requirement could be reduced through the application of state space estimation methods~\citep{erichsonShallowNeuralNetworks2020, fukamiGraspingExtremeAerodynamics2023}.

To approach zero-shot deployment, where the policy trained in simulation is applied directly to the physical wind farm, robust domain generalization strategies are essential.
These include training across a wide ensemble of inflow conditions (e.g., wind speed, turbulence intensity, thermal stratification).
Transfer to real wind farms may also benefit from online adaptation mechanisms, such as policy fine-tuning.
These would require safe exploration mechanisms to ensure operational stability during learning.

Our analysis suggests further lines of inquiry for future research.

\begin{enumerate}
    \item Constrained controller design: In this study, an emphasis was placed on not constraining the possible behavior of the controller too much, in order to allow flexibility in potential control strategies. This comes at the cost of convergence speed and computational burden. The results of our analysis suggest that a more effective way to design dynamic wind farm controllers could be to first obtain optimal static angles using Bayesian optimization, and then constrain the controller to only switch between the (finitely many) static optima obtained either directly from optimization or symmetry considerations.
    \item Sensor placement: The hyperparameter sweeps conducted in this study suggest that sensor location and the number of sensors are crucial for obtaining good performance. A dedicated study analyzing and optimizing sensor placement would be beneficial. Additionally, including both biased and noisy measurements in the training, that replicate real sensor constraints, may be beneficial in transferring simulation trained controllers to real wind farm application.
    \item Mechanical stresses: In the study, the reward was chosen as the instantaneous power output of the wind farm, in order to simplify convergence. In reality, moving the turbines will incur a cost in terms of both energy and mechanical stresses, which should be included in the reward design.
    \item Reference power signal tracking: Although this study focused on maximizing wind farm power output, it may be beneficial for grid balancing to train the RL controller to track a reference power signal. This would be particularly valuable for understanding controller behavior when the wind farm needs to reduce its power outcome.
    \item Combined induction–yaw control: Adding additional control parameters to wind turbines may allow greater flexibility for the trained controller. This may yield greater performance gains over a wider set of conditions. Control of the induction could be included as an additional output of the controller or as blade pitch angle or the generator torque of the turbine.
\end{enumerate}

\section{Methods}\label{sec:method}
We couple high-fidelity simulations of a wind farm with an \emph{active closed-loop controller}, which periodically updates the yaw angles of the wind turbines, in order to maximize the combined power output.
LES is used to model the three-dimensional flow through a series of wind turbines operating in an atmospheric boundary layer.
Periodically, we allow a closed-loop active controller to interact with the LES solver.
The controller is represented as a multi-layer perceptron (MLP) and dynamically updates the yaw angles of each turbine as a function of the wind velocities in the vicinity of the turbines.
The controller parameters are optimized using an RL setup, in order to maximize the total mean farm power output of the controlled system.

Section~\ref{subsec:windfarm} outlines the wind farm configuration and imposed wind conditions.
Section~\ref{subsec:les} details the LES solver and turbine modeling within the flow environment.
The RL framework and the corresponding optimization problem are described in Section~\ref{sec:rl}, while Section~\ref{sec:bo} presents the BO approach used to find the static baseline for comparison.
Finally, Section~\ref{sec:smartsim} discusses the integration of the LES code with the RL controller and our solution that allows efficient coupling.

\subsection{Wind Farm Configuration}\label{subsec:windfarm}

In this study a simple wind farm configuration with three turbines aligned in the $x$-direction and separated by a distance of $5D$ is considered, where the diameter of the wind turbine rotors is $D = 126m$.
The mean stream-wise velocity of the incoming flow is in the $x$-direction and has a magnitude of $U_{\infty} = 7.5 ms^{-1}$ at the wind turbine hub height, $y_h = 90m$. In the current work, the computational domain is of size $l_x, l_y, l_z = 19D, 4D, 7D$ and is discretized with a spacing $D/10$ into $n_x, n_y, n_z = 193, 41, 72$ grid points.
This grid spacing was chosen to capture the main turbulent structures in the flow while also providing accurate results of the turbines far wake flow and power output. It is consistent with the grid spacing used in previous studies~\citep{calafLESWindTurbine2011, revazLargeEddySimulationWind2021, bempedelisDatadrivenOptimisationWind, jane-ippelBayesianOptimisationTwoTurbine2024}.

\subsection{Large-Eddy Simulation (LES)}\label{subsec:les}

Three-dimensional LES are performed using Winc3D\citep{deskosWInc3DNovelFramework2020}, the wind farm simulator of the finite-difference framework XCompact3D\citep{bartholomew2020xcompact3d}.
Winc3D is used to predict the aerodynamics of the flow around the turbines and to determine the resulting power output of each turbine in the wind farm.
The simulations are based on the incompressible, explicitly filtered Navier–Stokes equations:

\begin{equation}\label{eq:les-ns-2}
\dt{\fil{u_i}} + \frac{1}{2} \left( \fil{u_j} \dxi{u_i} + \dxj{\fil{u_i} \fil{u_j}} \right) = - \frac{1}{\rho} \dxi{\fil{p}} - \dxj{\tau_{ij}} + \nu \dxjj{\fil{u_i}} + \frac{F_i}{\rho}, \quad (i = 1, 2, 3) 
\end{equation}

\begin{equation}\label{eq:les-ns-1}
\dxj{\fil{u_j}} = 0
\;.
\end{equation}
Here, spatial filtering is denoted by $\fil{\cdot}$ and is applied to the pressure $p$ and the velocity components $u_i$, with the summation implied over the $j$ index.
Fluid density and kinematic viscosity are denoted by $\rho$ and $\nu$, respectively.
External forces applied to the fluid (including the turbine model) are represented by $F_i$, and $\tau_{ij}$ are the sub-filter scale stresses, defined as $\tau_{ij} = \fil{u_i u_j} - \fil{u_i}\fil{u_j}$.

These stresses are modeled using the standard Smagorinsky model~\citep{Smagorinsky1963}:
\begin{equation}\label{eq:smagorinsky}
\tau_{ij} = -2 (C_s \Delta)^2 \fil{S_{ij}} |\fil{S}|,
\end{equation}

where $\fil{S_{ij}}$ is the filtered strain-rate tensor and $|\fil{S}|$ its magnitude.
The filter width $\Delta$ and the Smagorinsky coefficient $C_s$ define the sub-filter scale viscosity.
Near-wall damping is applied to $C_s$ using the formulation of~\citet{masonStochasticBackscatterLargeeddy1992}, where:

\begin{equation}\label{eq:MTdamp}
C_s = \left( C_0^n + \left[ \kappa \frac{y + y_0}{\Delta} \right]^{-n} \right)^{-1/n},
\end{equation}

where $C_0 = 1.4$ is the Smagorinsky constant far from the wall, $\kappa = 0.4$ is the von Kármán constant, $n = 3$ the growth parameter, and $y_0$ the surface roughness height.

Winc3D operates on a Cartesian mesh and employs high-order compact finite-difference schemes for spatial differentiation, filtering, and interpolation.
Spatial derivatives are computed using sixth-order implicit schemes, while temporal integration is performed using a third-order Adams–Bashforth method.
These high-order compact schemes offer quasi-spectral accuracy while allowing for non-periodic boundary conditions, enabling efficient resolution of small-scale turbulence with reduced computational cost~\citep{Laizet2009} that enables the long simulations needed for training the RL. The incompressibility of the velocity field is ensured by solving the pressure Poisson equation in spectral space using three-dimensional fast Fourier transforms (FFTs) and the concept of the modified wave number. The pressure grid is half-staggered from the velocity grid to avoid spurious pressure perturbations.

The simplicity of the Cartesian mesh is exploited with a 2D domain decomposition strategy implemented using standardized MPI that gives the code excellent strong and weak scaling properties \citep{Laizet2011, bartholomew2020xcompact3d}. The computational domain is divided into ``pencils'', each of which is handled by an MPI process.  The derivatives, interpolations, and one-dimensional FFTs in the x, y and z directions are performed within the X, Y, and Z pencils, allowing an efficient computation. Each operation is performed sequentially in one direction, with global transpositions enabling to switch from one pencil to another. 

\subsubsection{Actuator Disk Model}
Rather than solving the full geometry of the wind turbines, their effect on the wind can be modeled using the actuator surface~\citep{shenActuatorSurfaceModel}, actuator line~\citep{sorensenNumericalModelingWind2002} or actuator disc~\citep{mikkelsenActuatorDiscMethods} methods.
This work uses the actuator disc model because of its relatively lower computational cost.
The actuator disc model does not resolve the geometry of the individual turbine blades but instead uses a porous disc representation over their swept area.
Although small-scale flow structures are not captured in the wake, the work of~\citet{revazLargeEddySimulationWind2021} showed that the actuator disc model can provide accurate results of the far wake flow, as well as wind farm thrust and power predictions.
More detailed actuator line or surface models may offer better resolution in the near wake but considerably increase the simulation cost.

The implementation of the actuator disk model used here is detailed and validated in~\citet{bempedelisTurbulentEntrainmentFinitelength2023}, based on the method of~\citet{calafLargeEddySimulation2010} using the 1D momentum theory.
A uniform drag force is imparted on the flow over the swept area of the wind turbine blades, $A$, defined by,
\begin{equation}\label{eq:adm}
        F_t = -\frac{1}{2} \, \rho \, A \, C_T \left<\frac{U_r}{(1-a)}\right> ^2 \;,
\end{equation}
where $\left<\mathellipsis\right>$ denotes an averaging over the actuator disc and $U_r$ is the rotor normal velocity.
The axial induction factor, $a$, is derived from the thrust coefficient, $C_T$, such that $a= \frac{1}{2}\left(1 - \sqrt{1-C_T}\right)$.
This force is distributed over the disc area with a super-Gaussian smoothing applied, as described in~\citet{kingOptimizationWindPlant2017}.
The power output from the turbine can then be calculated as,
\begin{equation}\label{eq:admpower}
        P = -F_t U_r \;.
\end{equation}

\subsubsection{Atmospheric Boundary Layer (ABL)}\label{subsubsec:abl}
To generate the inflow conditions for the LES wind farm environments, precursor simulations are conducted to generate atmospheric boundary layers.
The atmospheric boundary layer is generated using a friction velocity $u^* = 0.442 ms^{-1}$, a boundary layer height $\delta = 504 m$, and a roughness length $z_0 = 0.05 m$.
These parameters were chosen to match those used in previous studies in~\citet{bempedelisDatadrivenOptimisationWind,wuModelingTurbineWakes2015} that best match the wind speed and turbulence intensity conditions observed and reported by~\citet{barthelmieModellingMeasuringFlow2009} that are representative of offshore wind conditions.
At each simulation step, a 2D slice of the flow field, orientated normal to the stream-wise direction, is stored to be used as the inflow condition for the wind farm simulations.
To minimize spurious low-frequency periodicity introduced by the recycling method in the precursor simulation, the domain was extended to a physical length of $224 D = 28,224 m$.
This helps to avoid artificial peaks in the frequency of the inflow caused by repeated large-scale structures.
Additionally, at each recycling step, the inflow plane is laterally shifted by one-eighth of the domain width to de-correlate repeating features~\citep{muntersShiftedPeriodicBoundary2016}.
These modifications were necessary to prevent the RL controller from exhibiting phase-locked behavior in response to periodic artifacts in the inlet when using shorter precursor domains, ensuring instead that it responds to the broadband turbulent features characteristic of the atmospheric boundary layer.

\subsection{Reinforcement Learning} \label{sec:rl}
RL is a machine learning approach that enables a controller (referred to as an agent in the RL literature) to optimize decision making by interacting with an environment, guided by a reward signal. In this section, we explain how we cast the wind farm optimization problem within the RL framework. We first outline the fundamental concepts of RL.

\subsubsection{Fundamentals of RL} \label{sec:rlfunadementals}
In RL, the environment is typically modeled as a discrete-time Markov decision process~\citep{sutton2018reinforcement}, defined by the tuple $(\mathcal{S}, \mathcal{A}, \mathcal{P}, \mathcal{R})$. Here, $\mathcal{S}$ denotes the set of states representing possible configurations of the system, $\mathcal{A}$ is the set of admissible actions available to the agent, and $\mathcal{P}(s' \mid s, a)$ is the state transition probability function, which describes the likelihood of transitioning to state $s'$ given that the agent takes action $a$ in state $s$. The reward function, $\mathcal{R}(s, a)$, assigns an immediate reward based on the chosen action and the resulting state.

A \emph{policy} $\pi: \mathcal{S} \rightarrow \mathcal{A}$ prescribes the agent's decision-making strategy by mapping states to actions. More generally, a stochastic policy defines a probability distribution over actions, $\pi(a \mid s)$, from which the agent samples its next move. Starting from an initial state $s_0$, the agent sequentially selects actions based on the policy, generating a trajectory or \emph{rollout} $\bar{s}_\pi = (s_0, a_0, s_1, a_1, r_1, \dots , s_k, a_k, r_k)$, where:
\begin{equation} 
a_k \sim \pi(\cdot \mid s_k ), \quad r_{k} = \mathcal{R}(s_k, a_k), \quad s_{k+1} \sim \mathcal{P}( \cdot \mid s_k, a_k). 
\end{equation}
The performance of a given policy is quantified by the \emph{return}, defined as the discounted sum of rewards along the trajectory: 
\begin{equation} \label{eq:return} R(\bar{s}_\pi) = \sum_{k=0}^L \gamma^k r_k, 
\end{equation} 
where $0 < \gamma \leq 1$ is the discount factor that determines the relative importance of immediate versus future rewards and $L > 0$ is the duration of an episode. The objective of RL is to identify the optimal policy, $\pi^*$, that maximizes the expected return: 
\begin{equation} \label{eq:optpolicycriterion}
\pi^* = \argmax_{\pi} \mathbb{E}_{s_0}[R_\pi]. 
\end{equation}
The role of the parameter $\gamma$ is to regularize the sum in equation~\eqref{eq:return}. Since $\sum_{k=0}^\infty \gamma^k = (1-\gamma)^{-1}$ whenever $0 < \gamma < 1$, it follows that the implication of choosing $\gamma < 1$ is to effectively truncate the policy return to a finite future time horizon of $(1-\gamma)^{-1}$ time steps. In practice, the $\argmax$ in equation~\eqref{eq:optpolicycriterion} is intractable, and instead the policy is parametrized as a deep neural network with parameters $\theta$, which are learned through gradient descent.

\subsubsection{Soft Actor Critic (SAC)}
Various RL algorithms have been developed to iteratively converge to an optimal policy as defined in equation~\eqref{eq:optpolicycriterion}.
Foundational methods include Q-Learning~\citep{watkins1989learning, watkins1992q}, policy iteration~\citep{howard1960dynamic}, temporal difference learning~\citep{sutton1988learning}, Double Q-Learning~\citep{van2016deep}, Deep Deterministic Policy Gradient~\citep{lillicrap2015continuous}, Twin Delayed Deep Deterministic Policy Gradient~\citep{fujimoto2018addressing} and Proximal Policy Optimization~\citep{schulman2017proximal}.
RL algorithms are notoriously sample inefficient, often requiring on the order of $10^5-10^7$ environment interactions to reach convergence~\citep{mnih2013playing, schulman2017proximal}.
Overestimation bias and the exploration-exploitation trade-off can lead to the discovery of suboptimal policies and lack of stability during training.
In addition, catastrophic forgetting can lead to policies which suddenly `unlearn' previously mastered tasks~\citep{zhang2022catastrophic, hafez2023map}.

Soft Actor Critic (SAC) is a state-of-the-art RL method for continuous control problems~\citep{haarnoja2018soft}.
SAC is a model-free RL algorithm without policies, which separates the policy (the actor) from the estimation of the expected return (the critic) as seen in Figure~\ref{fig:probes}(a), leading to improved stability of convergence and reducing the risk of overestimating the value of actions.
SAC extends traditional actor-critic methods by incorporating an entropy regularization term that balances the trade-off between exploration and exploitation into the objective.
The off-policy nature of SAC allows it to make use of an experience replay buffer, leading to higher sample efficiency compared to on-policy RL methods.
The off-policy nature of SAC in addition permits the use of parallel experience collection.
We simulate $M=32$ environments in parallel during training and $M=16$ environments during evaluation, significantly decreasing wall-clock time of the optimization loop.

Our implementation of SAC is based on the PyTorch machine learning library~\citep{paszke2019pytorch} and makes use of the TorchRL library for RL~\citep{bou2023torchrl}.

\subsubsection{States, Actions and Rewards for Wind Farm Optimization}
In order to cast our wind farm optimization problem in the RL framework, we define a discrete time step of $dt^{\text{RL}} = 10s$.
The LES solver interacts with the RL controller every $dt^{\text{RL}}$ time units, so that one discrete step of the RL rollout corresponds to $50$ discrete steps of the LES solver.
At each discretized time step, we update the yaw angles $\vec{\alpha} = (\alpha_1(t), \alpha_2(t), \alpha_3(t))$ of the individual turbines.
We employ a differential policy, which outputs the angular velocity with which each turbine moves.
This is done to ensure that the simulation remains numerically stable, as rapid changes in the yaw angle of the turbines are likely to lead to numerical issues and be impractical to perform with a real wind turbine.
The angles at time step $t$ are therefore updated as
\begin{equation}
\begin{gathered}
    \alpha_n(t) = \alpha_n(t-dt^{\text{RL}}) + v_n(t) \, dt^{\text{RL}}, \quad n=1,\dots,3, \\
    \vec{a}(t) = \vec{\alpha}(t) = \vec{\alpha}(t-dt^{\text{RL}}) + \vec{v}(t) \, dt^{\text{RL}}
\end{gathered}
\end{equation}
where $\vec{v}(t) = (v_1(t), v_2(t), v_3(t))$ is the vector of angular velocities of the turbines, which is the output of the policy $\pi$.
We limit the angular velocity to a maximum of $\left| v_n \right| \leq 1.0 \deg \, s^{-1}$.
In addition, we limit the maximum yaw angle to $\left| \alpha_n \right| \leq 40 \degree$ symmetrically.

The flow field is sensed through a total of $231$ velocity sensors arranged in a $2D$ plane at hub height surrounding the turbines, see Figure~\ref{fig:probes}(b).
The number of sensors was chosen as a result of a parameter sweep (see more details in the Appendix).
We only use the stream-wise velocity component $u_x$, which is the most informative (least noisy) of the three velocity components $(u_x, u_y, u_z)$.
The velocities of each individual sensor are averaged over the course of the discrete time step $dt^{\text{RL}}$ to smooth the signal.
The averaged sensor observations form the observation vector $\vec{o}(t)$.
We then define the state vector $\vec{s}(t)$ as
\begin{equation}
    \vec{s}(t) = (\vec{o}(t), \vec{\alpha}(t-dt^{\text{RL}})),
\end{equation}
where $\vec{\alpha}(t-dt^{\text{RL}})$ is the vector of turbine yaw angles from the previous time step. The state vector $\vec{s}(t)$ forms the input of the RL policy $\pi$.

\begin{figure}[htp]
    \begin{center}
    \includegraphics[valign=t,width=0.44\linewidth]{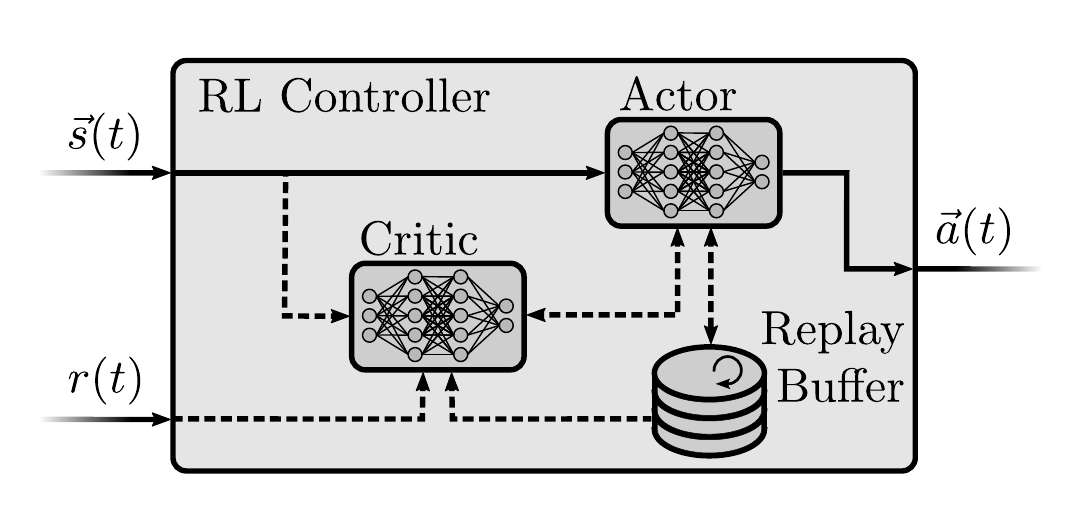}
    \includegraphics[valign=t,width=0.55\linewidth]{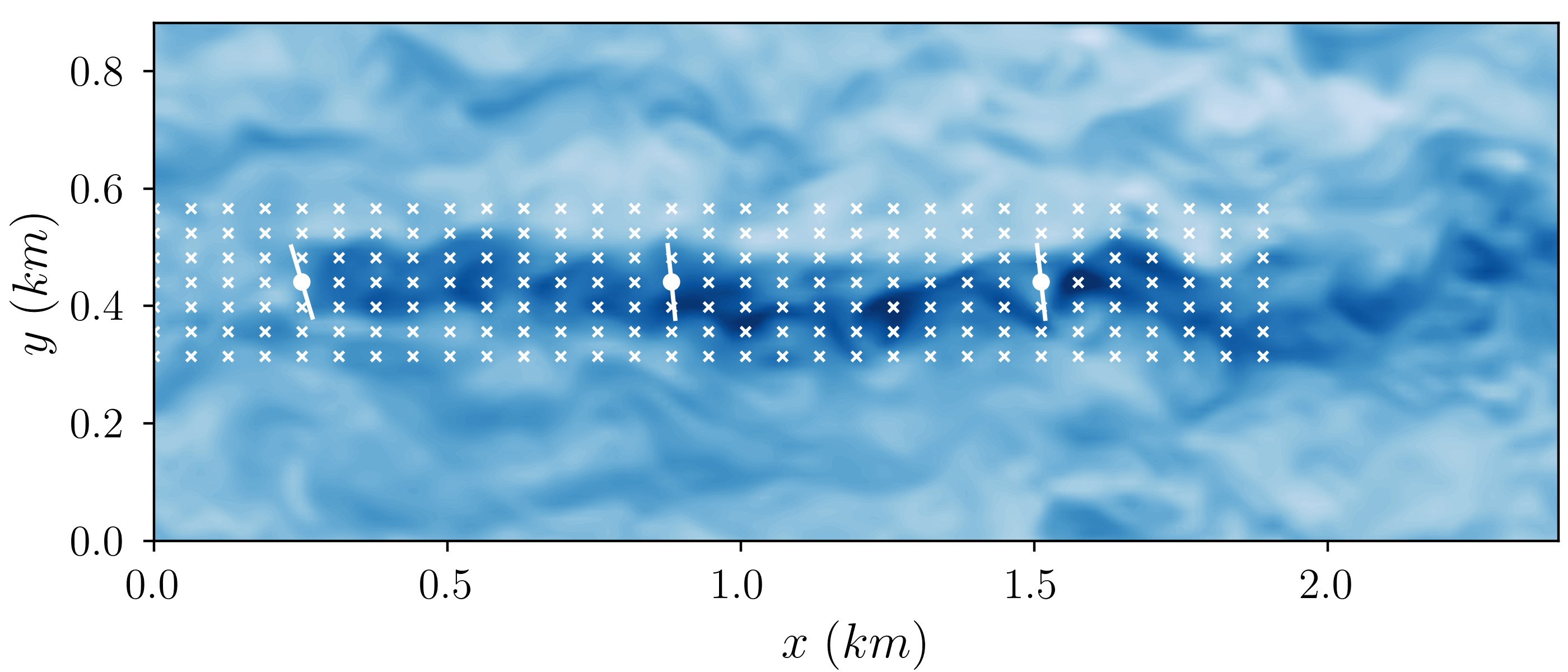}
    \caption{(a) Diagram showing the SAC controller algorithm with the dashed arrows showing the flow of information during training only and the solid arrows during training and evaluation. (b) Location of the velocity sensors around the turbines shown in an $x-z$ slice at the turbine hub height. The sensor locations are shown by the white crosses. The observations include the stream-wise velocity component as input to the RL policy, the instantaneous field of which is shown by the contours.}
        \label{fig:probes}
    \end{center}
\end{figure}

The reward is mainly composed of the instantaneous total wind farm power output.
The LES solver provides the instantaneous power output of each individual turbine, denoted as $P_n(t)$ for the $n$-th turbine for $n=1, \dots, 3$.
The instantaneous total power output of the wind farm can then be calculated as
\begin{equation} \label{eq:instpower}
     P(t) = \sum_{n=1}^3 P_n(t).
\end{equation}
In addition, we add a large angle penalty term, which prevents the controller from converging to very large yaw angles:
\begin{equation}
    r(t) = P(t) - \frac{\lambda}{3} \sum_{n=1}^3 \left( \frac{\alpha_n(t)}{\alpha_{\mathrm{max}}} \right)^{\kappa},
\end{equation}
where $P$ is defined in equation~\eqref{eq:instpower} and $\lambda$ and $\kappa$ are hyperparameters; see Table~\ref{table:rlparams}.
By integrating~\eqref{eq:instpower} in time, we can obtain the mean wind farm power output over time,
\begin{equation} \label{eq:meanfarmpower}
    \bar{P} = \lim_{T \rightarrow \infty} \frac{1}{T} \int_0^T P(t) \, dt.
\end{equation}
We note that, if $\displaystyle \gamma = 1$ and $\lambda = 0$, then $\displaystyle \lim_{L \rightarrow \infty} \frac{1}{L} \sum_{k=0}^L \gamma^k r_k = \bar{P}$.
Therefore, for this choice of hyperparameters, maximizing the return $R_\pi$ for a sufficiently long rollout is equivalent to maximizing the mean farm power $\bar{P}$. In practice, we use $\gamma < 1$, making the return a regularized (or truncated) version of the mean farm power.

\subsection{Bayesian Optimization} \label{sec:bo}
To establish a static baseline for comparison with the RL controller, we also perform BO.
In contrast to RL, which frames the control as a Markov decision process and learns a dynamic policy based on the observed system states over a series of time steps, BO identifies fixed yaw angles that maximize the mean wind farm power output:
\begin{equation}\label{eq:bo}
        \argmax_{\vec{\alpha}} \sum_{n=1}^N \overline{P}(\vec{\alpha})
        \quad
\text{subject to}
        \quad
    \vec{\alpha} \in [-\alpha_b, \; \alpha_b]
    \;.
\end{equation}
Bayesian optimization is an efficient sequential design strategy for global optimization of expensive to evaluate functions that uses a surrogate model to guide the search~\citep{jonesEfficientGlobalOptimization, NIPS2012_05311655, shahriariTakingHumanOut2016}.
The model is iteratively updated by adding new data at locations defined by an acquisition function.

In this work, the optimization is carried out using a Gaussian process surrogate model~\citep{rasmussenGaussianProcessesMachine2008} with a Matérn 5/2 kernel and automatic relevance determination and an upper confidence bound (UCB)~\citep{srinivasGaussianProcessOptimization} acquisition function implemented with BoTorch~\citep{balandat2020botorchframeworkefficientmontecarlo}.
The UCB exploration parameter is initialized at $\beta = 4$ and annealed linearly throughout the optimization.
The search space covers the yaw angles of all three turbines, with bounds defined by $\alpha_b = 40$ for each.
The optimization is initialized with 6 Latin hypercube samples and proceeds for 50 BO iterations.
Each optimization iteration consists of an ensemble of 32 LES, each averaged over 5,000 seconds of simulated time to produce $\overline{P}$.

\subsection{Coupling the RL controller with LES solver} \label{sec:smartsim}
Coupling the high-fidelity LES solver described in Section~\ref{subsec:les} with the RL controller described in Section~\ref{sec:rl} is challenging due to the different programming languages used.
While the LES solver is based on XCompact3D and is implemented in modern Fortran, the RL controller is implemented in Python using the PyTorch machine learning library.
At every RL time step (corresponding to $50$ LES solver steps), the controller processes the sensor observations computed by the LES solver, outputs updated yaw angles and communicates these updated angles to the LES solver.
This necessitates both the solver and the RL controller to frequently `wait' for the corresponding other code to compute a result before continuing.

We integrate the LES solver and RL controller components through the SmartSim library~\citep{smartsim}, an infrastructure framework that enables the integration of machine learning workflows into traditional HPC workloads.
In the SmartSim setup, the solver and controller are executed asynchronously on separate nodes and communicate with each other using SmartRedis through an additional orchestration and database node.
This approach is similar to previous work combing turbulent flow control with RL on HPC~\citep{fontDeepReinforcementLearning2025}.
We give a pseudocode outline of the system in Figure~\ref{fig:smartsimpseudo}.

\begin{figure}
    \centering
    \begin{minipage}[t]{0.48\textwidth}
        \centering
        \begin{algorithm}[H]
        \caption{LES solver (Fortran)}
        \begin{algorithmic}
        \State Initialize solver
        \While{training}
            \State \textbf{write} $\vec{s}$ and $P$ to database
            \While{controller not done}
                \State \textbf{wait}
            \EndWhile
            \State \textbf{read} $\vec{\alpha}$ from database
            \State \textbf{solve} $t \rightarrow t+1$ and  \textbf{update} $\vec{s}, P$
        \EndWhile
        \end{algorithmic}
        \end{algorithm}
    \end{minipage}
    \hfill
    \begin{minipage}[t]{0.48\textwidth}
        \centering
        \begin{algorithm}[H]
        \caption{RL controller (Python)}
        \begin{algorithmic}
        \State Initialize $\vec{\alpha} \gets 0$
        \While{training}
            \State \textbf{write} $\vec{\alpha}$ to database
            \While{solver not done}
                \State \textbf{wait}
            \EndWhile
            \State \textbf{read} $\vec{s}$ and $P$ from database
            \State $\vec{\alpha} \gets \pi_\theta(\vec{s})$
        \EndWhile
        \end{algorithmic}
        \end{algorithm}
    \end{minipage}
    \vspace{5mm}
    \caption{Communication loop for the LES solver (\emph{left}) and RL controller (\emph{right}). Here $\vec{s}$ denotes the state (velocity sensors and previous yaw angles) and $P$ denotes the wind farm power. Both codes are executed asynchronously on separate nodes and communicate with a separate database node. Synchronization is achieved by waiting for the respective counterpart to update a certain entry in the database. We have omitted the training part of the RL code to highlight the communication loop. During training, the entire loop is paused, and the parameters of the policy $\pi_\theta$ are updated. In practice, we use $M=32$ parallel instances of the LES environment during training and $M=16$ parallel instances during evaluation. Each instance is run on a single node of ARCHER2 (128 CPU cores).}
    \label{fig:smartsimpseudo}
\end{figure}

\section{Code Availability}
The LES code XCompact3d used for the simulations is available at \url{https://github.com/xcompact3d/Incompact3d}.
The code responsible for the RL and coupling with the LES is available at \url{https://github.com/admole/Wind-RL}.

\bibliography{Paper.bib}

\begin{thebibliography}{76}
\providecommand{\natexlab}[1]{#1}
\providecommand{\url}[1]{{#1}}
\providecommand{\urlprefix}{URL }
\providecommand{\doi}[1]{\url{https://doi.org/#1}}
\providecommand{\eprint}[2][]{\url{#2}}
 \bibcommenthead

\bibitem[{Balandat et~al(2020)Balandat, Karrer, Jiang, Daulton, Letham, Wilson,
  and Bakshy}]{balandat2020botorchframeworkefficientmontecarlo}
Balandat M, Karrer B, Jiang DR, et~al (2020) Botorch: A framework for efficient
  monte-carlo bayesian optimization.
  \urlprefix\url{https://arxiv.org/abs/1910.06403}, \eprint{1910.06403}

\bibitem[{Barthelmie and Jensen(2010)}]{barthelmieEvaluationWindFarm2010}
Barthelmie RJ, Jensen LE (2010) Evaluation of wind farm efficiency and wind
  turbine wakes at the {{Nysted}} offshore wind farm. Wind Energy
  13(6):573--586

\bibitem[{Barthelmie et~al(2009)Barthelmie, Hansen, Frandsen, Rathmann,
  Schepers, Schlez, Phillips, Rados, Zervos, Politis, and
  Chaviaropoulos}]{barthelmieModellingMeasuringFlow2009}
Barthelmie RJ, Hansen K, Frandsen ST, et~al (2009) Modelling and measuring flow
  and wind turbine wakes in large wind farms offshore. Wind Energy
  12(5):431--444

\bibitem[{Bartholomew et~al(2020)Bartholomew, Deskos, Frantz, Schuch,
  Lamballais, and Laizet}]{bartholomew2020xcompact3d}
Bartholomew P, Deskos G, Frantz R, et~al (2020) Xcompact3d: An open-source
  framework for solving turbulence problems on a cartesian mesh. SoftwareX
  12:100550

\bibitem[{Bastankhah and
  {Port{\'e}-Agel}(2014)}]{bastankhahNewAnalyticalModel2014}
Bastankhah M, {Port{\'e}-Agel} F (2014) A new analytical model for wind-turbine
  wakes. Renewable Energy 70:116--123

\bibitem[{Beckett et~al(2024)Beckett, Beech-Brandt, Leach, Payne, Simpson,
  Smith, Turner, and Whiting}]{archer2}
Beckett G, Beech-Brandt J, Leach K, et~al (2024) Archer2 service description.
  \urlprefix\url{https://doi.org/10.5281/zenodo.14507040}

\bibitem[{Bempedelis et~al(2023)Bempedelis, Laizet, and
  Deskos}]{bempedelisTurbulentEntrainmentFinitelength2023}
Bempedelis N, Laizet S, Deskos G (2023) Turbulent entrainment in finite-length
  wind farms. Journal of Fluid Mechanics 955:A12

\bibitem[{Bempedelis et~al(2024)Bempedelis, Gori, Wynn, Laizet, and
  Magri}]{bempedelisDatadrivenOptimisationWind}
Bempedelis N, Gori F, Wynn A, et~al (2024) Data-driven optimisation of wind
  farm layout and wake steering with large-eddy simulations. Wind Energy
  Science 9(4):869--882

\bibitem[{Bou et~al(2023)Bou, Bettini, Dittert, Kumar, Sodhani, Yang,
  Fabritiis, and Moens}]{bou2023torchrl}
Bou A, Bettini M, Dittert S, et~al (2023) Torchrl: A data-driven
  decision-making library for pytorch. \eprint{2306.00577}

\bibitem[{Bui et~al(2020)Bui, Nguyen, and
  Kim}]{buiDistributedOperationWind2020}
Bui VH, Nguyen TT, Kim HM (2020) Distributed {{Operation}} of {{Wind Farm}} for
  {{Maximizing Output Power}}: {{A Multi-Agent Deep Reinforcement Learning
  Approach}} 8

\bibitem[{Calaf et~al(2010)Calaf, Meneveau, and
  Meyers}]{calafLargeEddySimulation2010}
Calaf M, Meneveau C, Meyers J (2010) Large eddy simulation study of fully
  developed wind-turbine array boundary layers. Physics of Fluids 22(1):015110

\bibitem[{Calaf et~al(2011)Calaf, Meneveau, and
  Parlange}]{calafLESWindTurbine2011}
Calaf M, Meneveau C, Parlange M (2011) Large eddy simulation study of a fully
  developed thermal wind-turbine array boundary layer. In: Kuerten H, Geurts B,
  Armenio V, et~al (eds) Direct and Large-Eddy Simulation VIII. Springer
  Netherlands, Dordrecht, pp 239--244

\bibitem[{Chatzimanolakis et~al(2024)Chatzimanolakis, Weber, and
  Koumoutsakos}]{chatzimanolakis2024learning}
Chatzimanolakis M, Weber P, Koumoutsakos P (2024) Learning in two dimensions
  and controlling in three: Generalizable drag reduction strategies for flows
  past circular cylinders through deep reinforcement learning. Physical Review
  Fluids 9(4):043902

\bibitem[{Degrave et~al(2022)Degrave, Felici, Buchli, Neunert, Tracey,
  Carpanese, Ewalds, Hafner, Abdolmaleki, De~Las~Casas, Donner, Fritz,
  Galperti, Huber, Keeling, Tsimpoukelli, Kay, Merle, Moret, Noury, Pesamosca,
  Pfau, Sauter, Sommariva, Coda, Duval, Fasoli, Kohli, Kavukcuoglu, Hassabis,
  and Riedmiller}]{degraveMagneticControlTokamak2022}
Degrave J, Felici F, Buchli J, et~al (2022) Magnetic control of tokamak plasmas
  through deep reinforcement learning. Nature 602(7897):414--419

\bibitem[{Deskos et~al(2020)Deskos, Laizet, and
  Palacios}]{deskosWInc3DNovelFramework2020}
Deskos G, Laizet S, Palacios R (2020) {{WInc3D}}: {{A}} novel framework for
  turbulence-resolving simulations of wind farm wake interactions. Wind Energy

\bibitem[{Dong et~al(2021)Dong, Zhang, and Zhao}]{dongIntelligentWindFarm2021}
Dong H, Zhang J, Zhao X (2021) Intelligent wind farm control via deep
  reinforcement learning and high-fidelity simulations. Applied Energy
  292:116928

\bibitem[{Erichson et~al(2020)Erichson, Mathelin, Yao, Brunton, Mahoney, and
  Kutz}]{erichsonShallowNeuralNetworks2020}
Erichson NB, Mathelin L, Yao Z, et~al (2020) Shallow neural networks for fluid
  flow reconstruction with limited sensors. Proceedings of the Royal Society A:
  Mathematical, Physical and Engineering Sciences 476(2238):20200097

\bibitem[{Font et~al(2025{\natexlab{a}})Font, Alc{\'a}ntara-{\'A}vila, Rabault,
  Vinuesa, and Lehmkuhl}]{font2025deep}
Font B, Alc{\'a}ntara-{\'A}vila F, Rabault J, et~al (2025{\natexlab{a}}) Deep
  reinforcement learning for active flow control in a turbulent separation
  bubble. Nature Communications 16(1):1422

\bibitem[{Font et~al(2025{\natexlab{b}})Font, {Alc{\'a}ntara-{\'A}vila},
  Rabault, Vinuesa, and Lehmkuhl}]{fontDeepReinforcementLearning2025}
Font B, {Alc{\'a}ntara-{\'A}vila} F, Rabault J, et~al (2025{\natexlab{b}}) Deep
  reinforcement learning for active flow control in a turbulent separation
  bubble. Nature Communications 16(1):1422

\bibitem[{Frederik et~al(2020)Frederik, Doekemeijer, Mulders, and
  Van~Wingerden}]{frederikHelixApproachUsing2020}
Frederik JA, Doekemeijer BM, Mulders SP, et~al (2020) The helix approach:
  {{Using}} dynamic individual pitch control to enhance wake mixing in wind
  farms. Wind Energy 23(8):1739--1751

\bibitem[{Fujimoto et~al(2018)Fujimoto, Hoof, and
  Meger}]{fujimoto2018addressing}
Fujimoto S, Hoof H, Meger D (2018) Addressing function approximation error in
  actor-critic methods. In: International conference on machine learning, PMLR,
  pp 1587--1596

\bibitem[{Fukami and Taira(2023)}]{fukamiGraspingExtremeAerodynamics2023}
Fukami K, Taira K (2023) Grasping extreme aerodynamics on a low-dimensional
  manifold. Nature Communications 14(1):6480

\bibitem[{Goit and Meyers(2015)}]{goitOptimalControlEnergy2015}
Goit JP, Meyers J (2015) Optimal control of energy extraction in wind-farm
  boundary layers. Journal of Fluid Mechanics 768:5--50

\bibitem[{Haarnoja et~al(2018)Haarnoja, Zhou, Abbeel, and
  Levine}]{haarnoja2018soft}
Haarnoja T, Zhou A, Abbeel P, et~al (2018) Soft actor-critic: Off-policy
  maximum entropy deep reinforcement learning with a stochastic actor. In:
  International conference on machine learning, PMLR, pp 1861--1870

\bibitem[{Hafez et~al(2023)Hafez, Immisch, Weber, and Wermter}]{hafez2023map}
Hafez MB, Immisch T, Weber T, et~al (2023) Map-based experience replay: a
  memory-efficient solution to catastrophic forgetting in reinforcement
  learning. Frontiers in Neurorobotics 17:1127642

\bibitem[{{Harrison-Atlas} et~al(2024){Harrison-Atlas}, Glaws, King, and
  Lantz}]{harrison-atlasArtificialIntelligenceaidedWind2024a}
{Harrison-Atlas} D, Glaws A, King RN, et~al (2024) Artificial
  intelligence-aided wind plant optimization for nationwide evaluation of land
  use and economic benefits of wake steering. Nature Energy 9(6):735--749

\bibitem[{He et~al(2022)He, Zhao, Liang, Zhao, Qiu, and
  Dong}]{heEnsemblebasedDeepReinforcement2022}
He B, Zhao H, Liang G, et~al (2022) Ensemble-based {{Deep Reinforcement
  Learning}} for robust cooperative wind farm control. International Journal of
  Electrical Power \& Energy Systems 143:108406

\bibitem[{Houck(2022)}]{houckReviewWakeManagementTechniquies2022}
Houck DR (2022) Review of wake management techniques for wind turbines. Wind
  Energy 25(2):195--220

\bibitem[{Howard(1960)}]{howard1960dynamic}
Howard RA (1960) Dynamic programming and markov processes. MIT Press 2:39--47

\bibitem[{Howland et~al(2020)Howland, Ghate, Lele, and
  Dabiri}]{howlandOptimalClosedloopWake2020}
Howland MF, Ghate AS, Lele SK, et~al (2020) Optimal closed-loop wake steering
  -- {{Part}} 1: {{Conventionally}} neutral atmospheric boundary layer
  conditions. Wind Energy Science 5(4):1315--1338

\bibitem[{Howland et~al(2022)Howland, Quesada, Mart{\'i}nez, Larra{\~n}aga,
  Yadav, Chawla, Sivaram, and Dabiri}]{howlandCollectiveWindFarm2022}
Howland MF, Quesada JB, Mart{\'i}nez JJP, et~al (2022) Collective wind farm
  operation based on a predictive model increases utility-scale energy
  production. Nature Energy 7(9):818--827

\bibitem[{{International Energy Agency}(2021)}]{IEA2021}
{International Energy Agency} (2021) Net zero by 2050: A roadmap for the global
  energy sector. \urlprefix\url{https://www.iea.org/reports/net-zero-by-2050}

\bibitem[{Jan{\'e}-Ippel et~al(2024)Jan{\'e}-Ippel, Bempedelis, Palacios, and
  Laizet}]{jane-ippelBayesianOptimisationTwoTurbine2024}
Jan{\'e}-Ippel C, Bempedelis N, Palacios R, et~al (2024) Bayesian
  {{Optimisation}} of a {{Two}}-{{Turbine Configuration Around}} a {{2D Hill
  Using Large Eddy Simulations}}. Wind Energy p e2946

\bibitem[{Jensen(1983)}]{jensen1983}
Jensen N (1983) A note on wind generator interaction. No. 2411 in Ris{\o}-M,
  Ris{\o} National Laboratory

\bibitem[{Jim{\'e}nez et~al(2010)Jim{\'e}nez, Crespo, and
  Migoya}]{jimenezApplicationTechniqueCharacterize2010a}
Jim{\'e}nez {\'A}, Crespo A, Migoya E (2010) Application of a {{LES}} technique
  to characterize the wake deflection of a wind turbine in yaw. Wind Energy
  13(6):559--572

\bibitem[{Jones et~al(1998)Jones, Schonlau, and
  Welch}]{jonesEfficientGlobalOptimization}
Jones DR, Schonlau M, Welch WJ (1998) Efficient {{Global Optimization}} of
  {{Expensive Black-Box Functions}}. Journal of Global optimization 13:455--492

\bibitem[{King et~al(2017)King, Dykes, Graf, and
  Hamlington}]{kingOptimizationWindPlant2017}
King RN, Dykes K, Graf P, et~al (2017) Optimization of wind plant layouts using
  an adjoint approach. Wind Energy Science 2(1):115--131

\bibitem[{Kober et~al(2013)Kober, Bagnell, and
  Peters}]{koberReinforcementLearningRobotics2013}
Kober J, Bagnell JA, Peters J (2013) Reinforcement learning in robotics: {{A}}
  survey. The International Journal of Robotics Research 32(11):1238--1274

\bibitem[{Korb et~al(2021)Korb, Asmuth, Stender, and
  Ivanell}]{korbExploringApplicationReinforcement2021}
Korb H, Asmuth H, Stender M, et~al (2021) Exploring the application of
  reinforcement learning to wind farm control. Journal of Physics: Conference
  Series 1934(1):012022

\bibitem[{Laizet and Lamballais(2009)}]{Laizet2009}
Laizet S, Lamballais E (2009) High-order compact schemes for incompressible
  flows: A simple and efficient method with quasi-spectral accuracy. Journal of
  Computational Physics 228:5989--6015

\bibitem[{Laizet and Li(2011)}]{Laizet2011}
Laizet S, Li N (2011) {Incompact3d: A powerful tool to tackle turbulence
  problems with up to $\mathcal{O} (10^5)$ computational cores}. International
  Journal for Numerical Methods in Fluids 67:1735--57

\bibitem[{Lillicrap(2015)}]{lillicrap2015continuous}
Lillicrap T (2015) Continuous control with deep reinforcement learning. arXiv
  preprint arXiv:150902971

\bibitem[{Mason and Thomson(1992)}]{masonStochasticBackscatterLargeeddy1992}
Mason PJ, Thomson DJ (1992) Stochastic backscatter in large-eddy simulations of
  boundary layers. Journal of Fluid Mechanics 242:51--78

\bibitem[{Meyers and Meneveau(2012)}]{meyersOptimalTurbineSpacing2012}
Meyers J, Meneveau C (2012) Optimal turbine spacing in fully developed wind
  farm boundary layers. Wind Energy 15(2):305--317

\bibitem[{Meyers et~al(2022)Meyers, Bottasso, Dykes, Fleming, Gebraad, Giebel,
  G{\"o}{\c c}men, and Van~Wingerden}]{meyersWindFarmFlow2022a}
Meyers J, Bottasso C, Dykes K, et~al (2022) Wind farm flow control: Prospects
  and challenges. Wind Energy Science 7(6):2271--2306

\bibitem[{Mikkelsen(2004)}]{mikkelsenActuatorDiscMethods}
Mikkelsen R (2004) Actuator {{Disc Methods Applied}} to {{Wind Turbines}}. PhD
  thesis, Technical University of Denmark

\bibitem[{Mnih(2013)}]{mnih2013playing}
Mnih V (2013) Playing atari with deep reinforcement learning. arXiv preprint
  arXiv:13125602

\bibitem[{Mnih et~al(2015)Mnih, Kavukcuoglu, Silver, Rusu, Veness, Bellemare,
  Graves, Riedmiller, Fidjeland, Ostrovski et~al}]{mnih2015human}
Mnih V, Kavukcuoglu K, Silver D, et~al (2015) Human-level control through deep
  reinforcement learning. nature 518(7540):529--533

\bibitem[{Mnih et~al(2016)Mnih, Badia, Mirza, Graves, Lillicrap, Harley,
  Silver, and Kavukcuoglu}]{mnih2016asynchronous}
Mnih V, Badia AP, Mirza M, et~al (2016) Asynchronous methods for deep
  reinforcement learning. In: International conference on machine learning,
  PmLR, pp 1928--1937

\bibitem[{Mole and Laizet(2024)}]{moleMultifidelityBayesianOptimisation2024}
Mole A, Laizet S (2024) Multi-fidelity {{Bayesian Optimisation}} of {{Wind Farm
  Wake Steering}} using {{Wake Models}} and {{Large Eddy Simulations}}. Flow,
  Turbulence and Combustion

\bibitem[{Monroc et~al(2025)Monroc, Bu{\v s}i{\'c}, Dubuc, and
  Zhu}]{monrocWFCRLMultiAgentReinforcement2025}
Monroc CB, Bu{\v s}i{\'c} A, Dubuc D, et~al (2025) {{WFCRL}}: {{A Multi-Agent
  Reinforcement Learning Benchmark}} for {{Wind Farm Control}}.
  \eprint{2501.13592}

\bibitem[{Munters and Meyers(2018)}]{muntersDynamicStrategiesYaw2018}
Munters W, Meyers J (2018) Dynamic {{Strategies}} for {{Yaw}} and {{Induction
  Control}} of {{Wind Farms Based}} on {{Large-Eddy Simulation}} and
  {{Optimization}}. Energies 11(1):177

\bibitem[{Munters et~al(2016)Munters, Meneveau, and
  Meyers}]{muntersShiftedPeriodicBoundary2016}
Munters W, Meneveau C, Meyers J (2016) Shifted periodic boundary conditions for
  simulations of wall-bounded turbulent flows. Physics of Fluids 28(2):025112

\bibitem[{Padullaparthi et~al(2022)Padullaparthi, Nagarathinam, Vasan, Menon,
  and Sudarsanam}]{padullaparthiFALCONFArmLevel2022}
Padullaparthi VR, Nagarathinam S, Vasan A, et~al (2022) {{FALCON- FArm Level
  CONtrol}} for wind turbines using multi-agent deep reinforcement learning.
  Renewable Energy 181:445--456

\bibitem[{Partee et~al(2022)Partee, Ellis, Rigazzi, Shao, Bachman, Marques, and
  Robbins}]{smartsim}
Partee S, Ellis M, Rigazzi A, et~al (2022) Using machine learning at scale in
  numerical simulations with smartsim: An application to ocean climate
  modeling. Journal of Computational Science 62:101707

\bibitem[{Paszke et~al(2019)Paszke, Gross, Massa, Lerer, Bradbury, Chanan,
  Killeen, Lin, Gimelshein, Antiga et~al}]{paszke2019pytorch}
Paszke A, Gross S, Massa F, et~al (2019) Pytorch: An imperative style,
  high-performance deep learning library. Advances in neural information
  processing systems 32

\bibitem[{Rasmussen and Williams(2008)}]{rasmussenGaussianProcessesMachine2008}
Rasmussen CE, Williams CKI (2008) Gaussian Processes for Machine Learning, 3rd
  edn. Adaptive Computation and Machine Learning, MIT Press, Cambridge, Mass.

\bibitem[{Revaz and {Port{\'e}-Agel}(2021)}]{revazLargeEddySimulationWind2021}
Revaz T, {Port{\'e}-Agel} F (2021) Large-{{Eddy Simulation}} of {{Wind Turbine
  Flows}}: {{A New Evaluation}} of {{Actuator Disk Models}}. Energies
  14(13):3745

\bibitem[{Schulman et~al(2017)Schulman, Wolski, Dhariwal, Radford, and
  Klimov}]{schulman2017proximal}
Schulman J, Wolski F, Dhariwal P, et~al (2017) Proximal policy optimization
  algorithms. arXiv preprint arXiv:170706347

\bibitem[{Shahriari et~al(2016)Shahriari, Swersky, Wang, Adams, and {de
  Freitas}}]{shahriariTakingHumanOut2016}
Shahriari B, Swersky K, Wang Z, et~al (2016) Taking the {{Human Out}} of the
  {{Loop}}: {{A Review}} of {{Bayesian Optimization}}. Proceedings of the IEEE
  104(1):148--175

\bibitem[{Shen and N{\o}rk(2007)}]{shenActuatorSurfaceModel}
Shen WZ, N{\o}rk J (2007) Actuator {{Surface Model}} for {{Wind Turbine Flow
  Computations}}. In: Proceedings of European Wind Energy Conference 2007

\bibitem[{Smagorinsky(1963)}]{Smagorinsky1963}
Smagorinsky J (1963) General ciculation experiments with the primitive equa-
  tions. Monthly Weather Review 91(3):99--164

\bibitem[{Snoek et~al(2012)Snoek, Larochelle, and Adams}]{NIPS2012_05311655}
Snoek J, Larochelle H, Adams RP (2012) Practical bayesian optimization of
  machine learning algorithms. In: Pereira F, Burges C, Bottou L, et~al (eds)
  Advances in Neural Information Processing Systems, vol~25. Curran Associates,
  Inc.

\bibitem[{S{\o}rensen and Shen(2002)}]{sorensenNumericalModelingWind2002}
S{\o}rensen JN, Shen WZ (2002) Numerical {{Modeling}} of {{Wind Turbine
  Wakes}}. Journal of Fluids Engineering 124(2):393--399

\bibitem[{Srinivas et~al(2009)Srinivas, Krause, Kakade, and
  Seeger}]{srinivasGaussianProcessOptimization}
Srinivas N, Krause A, Kakade S, et~al (2009) Gaussian {{Process Optimization}}
  in the {{Bandit Setting}}: {{No Regret}} and {{Experimental Design}}. arXiv
  preprint arXiv:09123995

\bibitem[{Sutton(1988)}]{sutton1988learning}
Sutton RS (1988) Learning to predict by the methods of temporal differences.
  Machine learning 3:9--44

\bibitem[{Sutton and Barto(2018)}]{sutton2018reinforcement}
Sutton RS, Barto AG (2018) Reinforcement learning: An introduction. MIT press

\bibitem[{Van~Hasselt et~al(2016)Van~Hasselt, Guez, and Silver}]{van2016deep}
Van~Hasselt H, Guez A, Silver D (2016) Deep reinforcement learning with double
  q-learning. In: Proceedings of the AAAI conference on artificial intelligence

\bibitem[{Veers et~al(2023)Veers, Bottasso, Manuel, Naughton, Pao, Paquette,
  Robertson, Robinson, Ananthan, Barlas, Bianchini, Bredmose, Horcas, Keller,
  Madsen, Manwell, Moriarty, Nolet, and
  Rinker}]{veersGrandChallengesDesign2023}
Veers P, Bottasso CL, Manuel L, et~al (2023) Grand challenges in the design,
  manufacture, and operation of future wind turbine systems. Wind Energy
  Science 8(7):1071--1131

\bibitem[{Vignon et~al(2023)Vignon, Rabault, and Vinuesa}]{vignon2023recent}
Vignon C, Rabault J, Vinuesa R (2023) Recent advances in applying deep
  reinforcement learning for flow control: Perspectives and future directions.
  Physics of Fluids 35(3)

\bibitem[{Wang et~al(2025)Wang, He, Yan, Han, and
  Liu}]{wangDeepReinforcementLearningdriven2025}
Wang H, He S, Yan J, et~al (2025) Deep reinforcement learning-driven wind farm
  flow control considering dynamic wind. Energy Conversion and Management
  337:119888

\bibitem[{Watkins and Dayan(1992)}]{watkins1992q}
Watkins CJ, Dayan P (1992) Q-learning. Machine learning 8:279--292

\bibitem[{Watkins(1989)}]{watkins1989learning}
Watkins CJCH (1989) Learning from delayed rewards

\bibitem[{Wu and {Port{\'e}-Agel}(2015)}]{wuModelingTurbineWakes2015}
Wu YT, {Port{\'e}-Agel} F (2015) Modeling turbine wakes and power losses within
  a wind farm using {{LES}}: {{An}} application to the {{Horns Rev}} offshore
  wind farm. Renewable Energy 75:945--955

\bibitem[{Xia et~al(2024)Xia, Zhang, Kerrigan, and Rigas}]{xia2023active}
Xia C, Zhang J, Kerrigan EC, et~al (2024) Active flow control for bluff body
  drag reduction using reinforcement learning with partial measurements.
  Journal of Fluid Mechanics 981:A17

\bibitem[{Zhang et~al(2022)Zhang, Wang, Liang, and
  Yuan}]{zhang2022catastrophic}
Zhang T, Wang X, Liang B, et~al (2022) Catastrophic interference in
  reinforcement learning: A solution based on context division and knowledge
  distillation. IEEE Transactions on Neural Networks and Learning Systems
  34(12):9925--9939

\end{thebibliography}

\section*{Acknowledgments}
The authors gratefully acknowledge the support of the UK Engineering and Physical Sciences Research Council (EPSRC) for funding this research through the AI for Net Zero grant ``Real-time digital optimisation and decision making  for energy and transport systems'' (EP/Y005619/1)
Access to the ARCHER2 UK National Supercomputing Service was provided through the UK Turbulence consortium (EP/R029326/1 and EP/X035484/1).
The authors would also like to thank Eleanor Broadway and Joseph Lee for their help in setting up the coupling on ARCHER2.

\section*{Author contributions}
A.M. and M.W. conceived the study, designed the methodology, implemented the control algorithms, integrated the RL and LES components, analyzed the results, created the figures used in the manuscript, wrote the first draft of the manuscript, revised and edited the manuscript.
A.M. ran the simulations, managed the HPC resources and simulation data.
G.R. and S.L. conceived the study, revised and edited the manuscript, provided project guidance and technical oversight, acquired funding and managed project resources.

\appendix

\section*{Appendix: Time Lag Correction}

\begin{figure}[H]
    \centering
    \includegraphics[width=\linewidth]{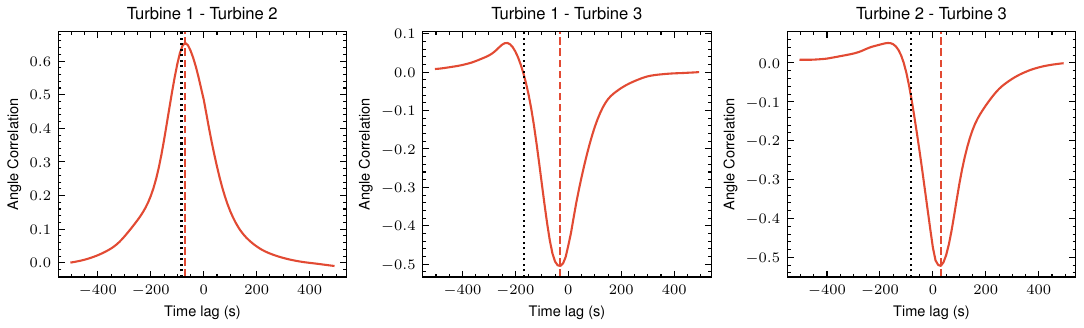}
    \includegraphics[width=\linewidth]{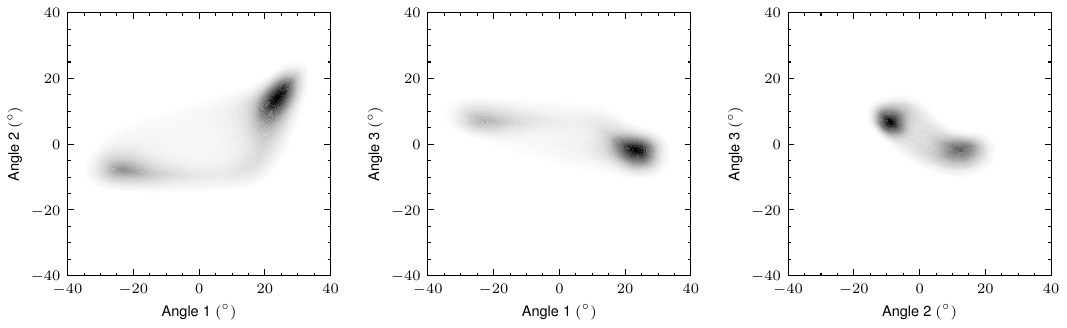}
    \includegraphics[width=\linewidth]{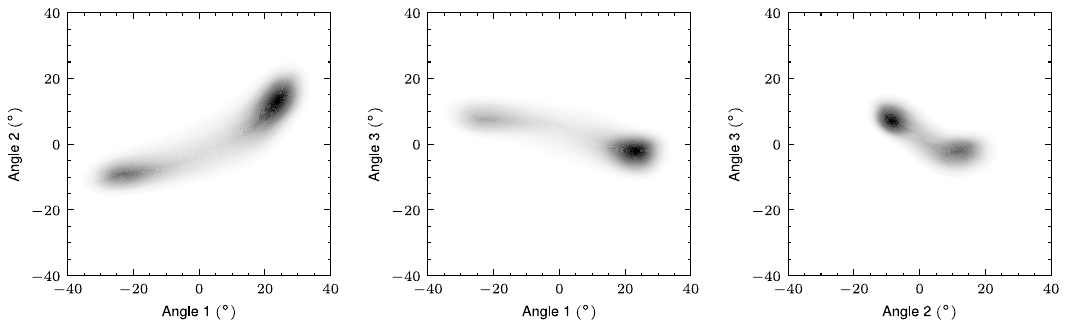}
    \caption{Cross-correlation analysis of yaw angle pairs. Kendall's tau cross-correlation functions are shown for \emph{(a)} Angle 1 – Angle 2, \emph{(b)} Angle 1 – Angle 3, and \emph{(c)} Angle 2 – Angle 3 pairs across a range of time lags. The red dashed vertical line indicates the lag corresponding to the peak cross-correlation, highlighting the characteristic response time between variables. The black dotted vertical line marks the expected advection time for each pair, providing a reference for physical transport processes. \emph{(d-f)} Joint probability density plots of angle pairs before correcting for time shifts and \emph{(g-i)} after correcting for time-shifts determined from the cross-correlation analysis. Each plot shows the distribution of observed angle combinations for the respective variable pairs. Time shift correction results in increased density along the diagonal, reflecting better synchronization between yaw angles.}
    \label{fig:angle_cross}
\end{figure}

\pagebreak
\section*{Appendix: Yaw Angle Correlation Linear Models}

The fitted models, shown in figure~\ref{fig:controller}(c)(d)(e) are given by:
\begin{align}
\alpha_2 &= (0.2747 \pm 0.0083) + (0.4492 \pm 0.0004)\,\alpha_1\,, \\
\alpha_3 &= (3.2572 \pm 0.0061) + (-0.2269 \pm 0.0003)\,\alpha_1\,, \\
\alpha_3 &= (3.1932 \pm 0.0061) + (-0.4256 \pm 0.0006)\,\alpha_2\,,
\end{align}
where we indicate the fitted parameters as $m \pm s$ with $m$ the estimated parameter and $s$ its estimated standard error.

\section*{Appendix: Barycenter Distribution}

\begin{figure}[H]
    \centering
    \includegraphics[width=\linewidth]{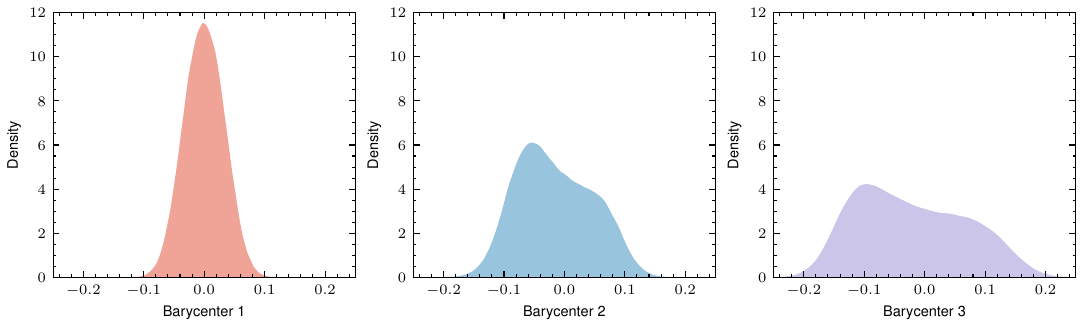}
    \caption{Probability density functions (PDFs) of the stream-wise velocity barycenter positions for three locations upstream of \emph{(a)} Turbine 1, \emph{(b)} Turbine 2 and \emph{(c)} Turbine 3. PDFs are calculated using kernel density estimation with a Gaussian kernel and bandwidth selected according to Scott’s rule. The barycenter position represents the spatial centroid of the velocity deficit region, providing a statistical description of upstream flow structures influencing turbine inflow. In front of Turbine 1 the flow is statistically symmetrical with asymmetry introduced downstream by the RL yaw control.}
    \label{fig:barycenter}
\end{figure}

\pagebreak

\section*{Appendix: Reinforcement Learning Hyperparameters}
\FloatBarrier
\vspace{-8mm}
\renewcommand{\arraystretch}{1.2}
\begin{table}[h]
	\caption{Values of the relevant SAC hyperparameters  and wind farm environment parameters. The column `Training' contains the values used for the final training runs. The column `Sweeps` contains either the standard value used, or the range of values considered in hyperparameter sweeps.}
     \label{table:rlparams}
	\begin{tabular}{@{} l @{\hspace{5em}} c @{\hspace{5em}} c @{}}
		\toprule
		Parameter & Training & Sweeps \\
		\midrule
            Discount factor $\gamma$ & $0.99$ & $[0.8, 0.99]$ \\
            Learning rate (actor and critic) & $3 \times 10^{-6}$ & $[3 \times 10^{-6}, 3 \times 10^{-4}]$ \\
            Learning rate (entropy) & $1 \times 10^{-5}$ & $[1 \times 10^{-6}, 3 \times 10^{-4}]$ \\
            Initial entropy loss multiplier & $10$ & $[1, 100]$ \\
            Optimizer batch size & $256$ & $256$ \\
            Frames per batch & $512$ & $512$ \\
            Number of probes per turbine & $77$ & $[10, 77]$ \\
            Initial random steps & $3.2 \times 10^5 s$ & $3.2 \times 10^5 s$ \\
            Episode length & $5 \times 10^3 s$ & $5 \times 10^3 s$ \\
            Number of episodes & $2000$ & $600$ \\
            Number of parallel environments & $32$ & $32$ \\
            \midrule
            Maximum yaw angle $\alpha_{\mathrm{max}}$ & $40$ deg & $40$ deg \\
            Maximum yaw velocity $v_{\mathrm{max}}$ & $1.0 \deg s^{-1}$ & $1.0 \deg s^{-1}$ \\
            $\lambda$ (reward parameter) & 0.5 & 0.5 \\
            $\kappa$ (reward parameter) & 26 & 26 \\
            $dt_{\mathrm{RL}}$ & $10s$ & $10s$ \\
            $dt_{\mathrm{LES}}$ & $0.2s$ & $0.2s$ \\
		\botrule
	\end{tabular}
\end{table}
\renewcommand{\arraystretch}{1.0}
\FloatBarrier

\begin{figure}[H]
    \centering
    \includegraphics[width=0.8\linewidth]{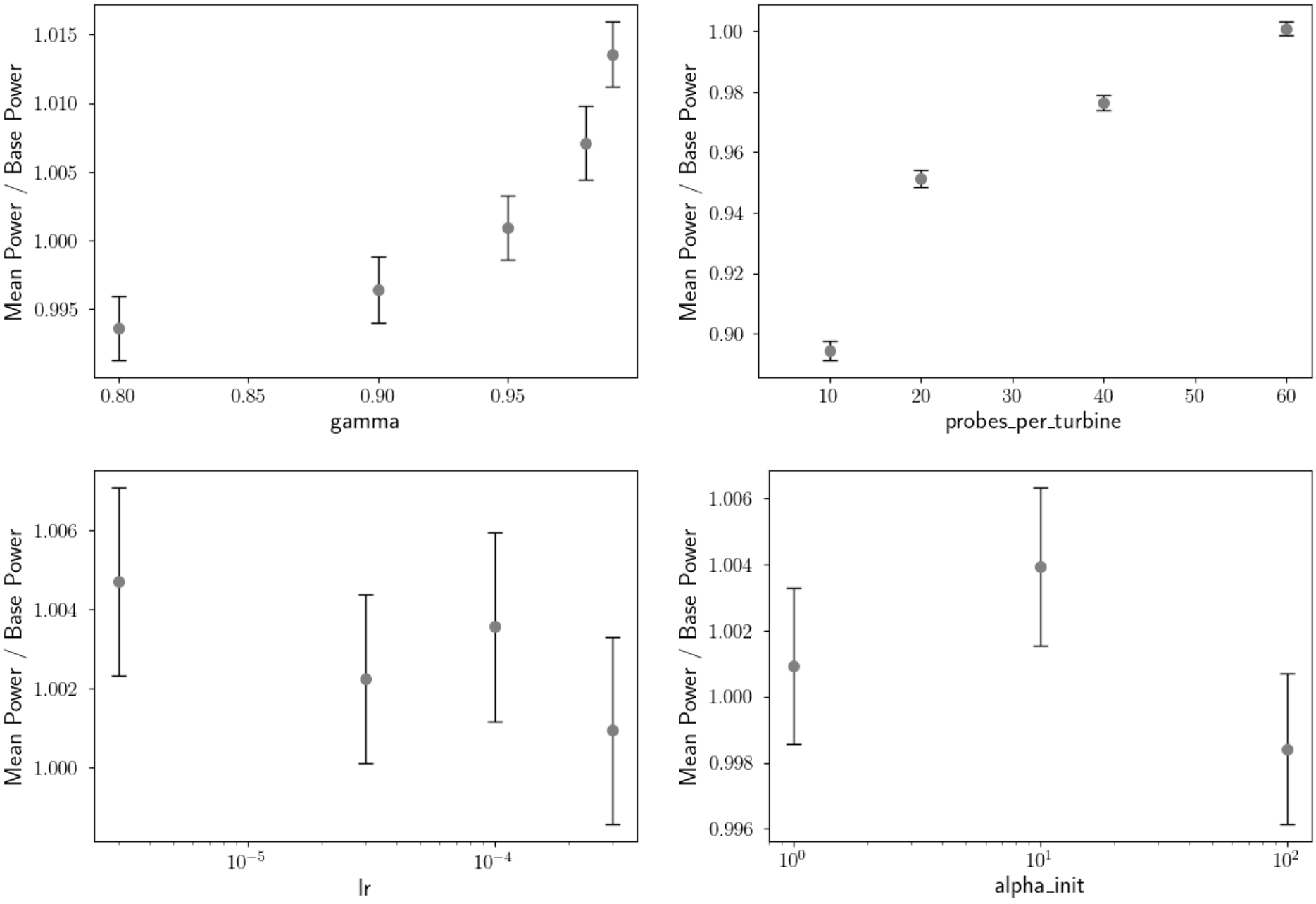}
    \caption{The impact on wind farm power output during evaluation of RL controllers trained with varying (a) discount factor (gamma), (b) number of probes per turbine, (c) learning rate for both the actor and critic networks, and (d) initial entropy loss multiplier (alpha init). Each panel presents the results of sweeping one parameter while holding others constant, with performance evaluated based on the rewards obtained from an evaluation of the trained controller.}
    \label{fig:sweep}
\end{figure}

\end{document}